\documentclass[12pt]{article}
\usepackage{amsmath,amssymb,cite,hyperref,tikz}
\usetikzlibrary{intersections}

\newcommand{\la}{\langle}
\def\cA{{\mathcal A}}
\def\cB{{\mathcal B}}

\def\cG{{\mathcal G}}
\def\cI{{\mathcal I}}
\def\cM{{\mathcal M}}
\def\cN{{\mathcal N}}
\def\cL{{\mathcal L}}

\newcommand{\cO}{\mathcal{O}}
\def\cX{{\mathcal X}}
\def\cZ{{\mathcal Z}}

\newcommand{\CP}{\mathbb{CP}}
\newcommand{\M}{\mathbb{M}}
\newcommand{\R}{\mathbb{R}}
\newcommand{\C}{\mathbb{C}}
\def\p{\partial}

\DeclareMathOperator{\Gr}{Gr}
\DeclareMathOperator{\tr}{tr}

\begin{document}

\thispagestyle{empty}

\null\vskip-43pt \hfill
\begin{minipage}[t]{50mm}
DCPT-16/59 \\
\end{minipage}

\vskip 2.2truecm
\begin{center}
\vskip 0.2truecm

 {\Large\bf
The Correlahedron
}
\vskip 0.5truecm

\vskip 1truecm
{\bf   Burkhard Eden$^{a}$, Paul Heslop$^{b}$,   Lionel Mason$^{c}$ \\
}

\vskip 0.4truecm
 $^{a}$  {\it Institut f\"ur Mathematik und Physik, Humboldt-Universit\"at zu Berlin,
 	\\
 	Zum grossen Windkanal 6, 12489 Berlin}\\
\vskip .2truecm
$^{b}$ {\it 
	Mathematics Department, Durham University,
	Science Laboratories,
	\\South Rd, Durham DH1 3LE}
 \\
 \vskip .2truecm
$^{c}$  {\it The Mathematical Institute, University of Oxford, AWB ROQ, Woodstock Road, OX2 6GG  } \\

\end{center}

\vskip 2truecm

\centerline{\bf Abstract} 
\medskip \noindent 
We introduce a new geometric object, the correlahedron, which we conjecture to be equivalent to stress-energy correlators in planar $\cN=4$ super Yang-Mills. Re-expressing the Grassmann dependence of  correlation functions of $n$ chiral stress-energy multiplets  with Grassmann degree $4k$ in terms of $4(n+k)$-linear bosonic variables, the resulting expressions have an interpretation as volume forms on a $\Gr(n{+}k,4{+}n{+}k)$ Grassmannian, analogous to the expressions for planar amplitudes via the amplituhedron. The resulting volume forms are to be naturally associated with the correlahedron geometry. We construct such expressions in this bosonised space both directly, in general, from Feynman diagrams in twistor space, and then more invariantly from specific known correlator expressions in analytic superspace. We give a geometric interpretation of the action of the consecutive lightlike limit and show that under this the correlahedron reduces to the squared amplituhedron both as a geometric object as well as directly on the corresponding volume forms. We give an explicit easily implementable algorithm via cylindrical decompositions for extracting the squared amplituhedron volume form from the squared amplituhedron geometry with explicit examples and discuss the analogous procedure for the correlators.

\newpage

\thispagestyle{empty}

{\small \tableofcontents}

\newpage
\setcounter{page}{1}\setcounter{footnote}{0}

\section{Introduction} 

Both scattering amplitudes and stress-tensor correlators in $\cN=4$ SYM have been
the subject of intense research for a
number of years, revealing  wonderful discoveries of
mathematical structures.  We will be focusing on the integrand in
this paper following much recent work (see for example~\cite{
 ArkaniHamed:2010kv,Boels:2010nw,Bullimore:2010pj, Mason:2010yk,
 Bullimore:2010dz,Eden:2010ce,Eden:2011ku,Bourjaily:2016evz} and
references therein). One of the most exciting discoveries is
that the perturbative integrands of
$n$-point, $\ell$-loop  scattering amplitudes in planar
$\cN=4$ SYM are equivalent to generalised  polyhedra in Grassmannians, 
with faces and vertices determined by 
the momenta and helicities of the particles being scattered~\cite{
Arkani-Hamed:2013jha}. This
geometrical object was named the {\em amplituhedron} (see also further developments in~\cite{Arkani-Hamed:2013kca,Arkani-Hamed:2014dca,Lam:2014jda,Bai:2014cna,Franco:2014csa,Bern:2015ple,Galloni:2016iuj,Ferro:2016zmx}).
On the other hand (the square of)  all $\ell$-loop amplitudes are limits of tree-level correlation
functions of the stress-energy
multiplet (correlators)~\cite{Alday:2010zy,Eden:2010zz,Eden:2010ce,Eden:2011yp,Adamo:2011dq,Eden:2011ku}
suggesting the possibility of a larger geometrical object
describing correlators and reducing to the amplituhedron in
relevant 
limits. The purpose of this paper is to give a proposal for this  {\em correlahedron}.

\vspace{.2cm}

The starting point for the amplituhedron was the introduction of momentum supertwistors followed by a ``bosonisation'' of their fermionic parts.  Hodges showed that they lead to a geometric formulation of the determinants arising from the fermionic coordinates as volumes  of a polyhedron in a projective space for the NMHV amplitude \cite{Hodges:2009hk}.
 To generalize\footnote{In fact this generalization is really that of the dual of the original Hodges framework~\cite{Arkani-Hamed:2014dca}.} to higher MHV degree introduce a 
particle-independent fermionic variable $\phi_I^p$, $p=1,\ldots,k$ where $I=1,\ldots , 4$ is an $R$-symmetry index, and send the odd variables $\chi_i^I$ to even variables 
$\xi_i^p= \chi_i^I \phi_I^p$~\cite{Arkani-Hamed:2013jha}. Here the range of the index
$p$ depends on the helicity structure (or Grassmann degree) of the superamplitude;  for
N${}^k$MHV amplitudes $p=1,\ldots,k$ and thus momentum
supertwistor space $(Z,\chi)$ becomes the vector space  $\C^{4+k}$
with bosonic variables $(Z,\xi)$. This framework has
considerable practical advantages -- for example nilpotent superconformal
invariants are straightforward to find and  non-trivial
superconformal identities become manifest generalized Schouten
identities. Furthermore the resulting expression can be seen to arise from volume forms on the Grassmannian of $k$-planes in $4{+}k$ dimensions,
$\Gr(k,4{+}k)$. The construction essentially reduces superconformal invariants to projective invariants.

We perform an analogous bosonisation of the stress-tensor
correlators. It is  not immediately clear how to do this 
bosonisation 
starting from supercorrelators in analytic super-space directly. However, recently
such correlators were considered via Feynman diagrams in supertwistor
space~\cite{Chicherin:2014uca} and this formulation leads to a ``potential'' for the correlation functions.  This potential is   a correlator of certain `log det d-bar' operators based on lines in twistor space. These operators are not manifestly gauge
invariant, but only become so when differentiated by a fourth order
Grassmann odd differential operator at each point mapping the `log det
d-bar' operators to 
the gauge invariant super-BPS operators $\cO_i$.
In the diagram formulation based on an axial gauge, the gauge
dependence will manifest itself in dependence on the reference twistor
$Z_*$.  We will nevertheless suppress this differentiation in the following and
indeed provide ample evidence for the conjecture that there is a $Z_*$ independent `potential'
of the sum of diagrams given by the correlahedron. Indeed there is a
simple prescription for lifting this
$Z_*$-independent potential directly from analytic superspace, even
though it is not 
obtained by direct bosonisation of the analytic superspace correlator.
We thus rewrite all known stress-tensor supercorrelators in an
appropriately bosonised form. These expressions are all equivalent to 
volume forms on the Grassmannian space $\Gr(k{+}n,4{+}k{+}n)$.

The key aspect of the amplituhedron however is geometric; it is a 
generalised polyhedron  lying in the \emph{real} Grassmannian
$\Gr(k,4{+}k)$.  A natural volume form on this polyhedron, one with log
divergences on the boundary and no divergences inside,  gives the afore
mentioned bosonised amplitudes.
 We generalise this geometric aspect to
the correlahedron, now lying in $\Gr(k{+}n,4{+}k{+}n)$. More precisely it is the
``squared amplituhedron'', 
a larger object than the amplituhedron itself  which
generalises to the correlahedron. This ``squared amplituhedron''
corresponds 
to the square of the superamplitude.  A key advantage of the squared amplituhedron is that it has a more 
explicit definition than the amplituhedron itself being simply defined by explicit inequalities, whereas the amplituhedron requires a further topological degree requirement~\cite{Arkani-Hamed:2017vfh}.

The lightlike limit, by which the correlators
become the square of superamplitudes, has a natural
geometrical interpretation for the correlahedron. Under a partial freezing to a
boundary of the correlahedron space, together with a projection,
the correlahedron geometry becomes the squared amplituhedron
geometry. This same procedure projects the corresponding correlator
volume form to the squared amplitude  volume form.

For the amplituhedron, the link between the integrands and the geometry  arises from
the requirement that the  volume form
should have no divergences inside the amplituhedron and log divergences on its boundary. This volume
form {\em is} essentially the bosonised amplitude. Obtaining this form from
the geometry is non-trivial for the amplituhedron, but becomes
much simpler for the ``squared amplituhedron'' due to its more
explicit definition.  A key point is that the requirement that the volume form have simple poles on the boundary is not sufficient to determine it, but the combinatorics of the positive  geometry of the polyhedral description does.  This is manifested in a by-product of this work in which we
introduce a completely  algorithmic and easily computerisable way of
obtaining this volume form from the geometry of the squared
amplituhedron. The algorithm uses 
cylindrical decomposition, an active area of research in its own
right and with a number of physical applications, which unfortunately
however can be doubly exponential in the number of variables. This
method quite quickly becomes impractical for large particle number
or loop order.  Nevertheless in a number of non-trivial
examples, we show that the squared amplitude geometry gives the square of
the superamplitude.
We then explore the corresponding relation between the correlahedron
and the bosonised correlators.

The plan of the paper is thus as follows.
In section~\ref{hedrons} we introduce our conventions, details of
the bosonisation procedure, and the definitions of the Grassmannians
in which the various 'hedra lie. In section~\ref{sec:hedron-geometry}
we then define the various hedra -- amplituhedron, squared
amplituhedron and correlahedron --  as geometrical polytopes in
the corresponding 
Grassmannians. In section~\ref{sec:hedron-expressions} we discuss
how to  write  known  
 explicit expressions for correlators as volume forms on
the appropriate Grassmannian. In
section~\ref{sec:lightlike-limit} we consider the lightlike limit of
correlators in correlahedron space. We show that the same geometric
procedure reduces the geometry of the correlahedron
to that of the amplituhedron as well as reducing the corresponding
volume form 
expressions to those of the amplituhedron volume form
expressions. Finally in section~\ref{sec:computations} we consider the
connection between the hedron geometry and the hedron volume forms. We
develop a simple algorithm using  cylindrical decomposition for
obtaining the volume form from the geometry and apply it to a number
of squared amplituhedrons and a correlahedron example.
In an appendix we look at the most non-trivial examples of taking the
lightlike  
limit of the correlahedron.

\section{Bosonisation, conventions and -hedron forms} \label{hedrons}

A key aspect of both the amplituhedron and correlahedron is the
bosonised superspace. As such in this section we review this procedure
for amplitudes and give our proposal for the appropriate bosonised
space for correlators. This will also set out our notation
and conventions for the rest of the paper.

\subsection{Bosonisation}
 
Planar superamplitudes in $\cN$=4 SYM can be nicely presented in  momentum
supertwistor space $\C^{4|4}$~\cite{Hodges:2009hk}. Bosonisation of superspace
for N${}^{k'}$MHV superamplitudes
maps momentum supertwistors, which lie in $(4|4)$ dimensions,  to a
purely bosonic vector of dimension $4+k'$:
\begin{align}
\mathbb{C}^{4|4}\ni  \quad 
 \left(
 z| \chi
 \right)\ \rightarrow \ Z=\left(
z, \zeta
\right) = \left(
z, \chi  \phi
  \right)\quad \in \mathbb{C}^{4+k'}\ .
\end{align}
Here $z$ is a bosonic four-dimensional row vector (a twistor), $\chi$
is a fermionic 4-vector (the Grassmann odd component of the
supertwistor) and $\phi$ is a Grassmann odd $4\times k'$ matrix. Thus
$Z$ is indeed a Grassmann even (bosonic) $4+k'$-dimensional row
vector.

The amplituhedron
space itself is a subset of $\Gr(k',k'+4)$, the space of $k'$-planes, $Y$, in
$4+k'$ dimensions.

The (chiral) correlator on the other hand will be written in terms of a
potential on
chiral superspace in section~\S\ref{potentials}. Chiral
super-Minkowski space can be  equivalently thought of as the space of 2-planes in supertwistor
space. Such 2-planes on supertwistor space are specified by taking two
independent supertwistors on the plane. We will write them as a
$2\times(4|4)$ supermatrix $(x|\theta)$ where the two rows of
the matrix are the two supertwistors in question, and there is a local
$GL(2)$ acting on the left, corresponding to the independence of the
plane on the choice of the two supertwistors.%
\footnote{This $GL(2)$ symmetry can be used to set the 2$\times$ 4
  matrix $x$ to the form $(1_2,\hat x)$ where $\hat x$ is a 2$\times$2
  matrix, the standard spinor representation of 4d Minkowski space.} 

We perform a very similar bosonisation of these co-ordinates as perfomed above
for the amplitudes, with the main difference now being that the bosonised
supertwistor space lives in $4+n+k$ dimensions rather than $k'+4$
dimensions.  So explicitly we map the $2\times(4|4)$ supermatrix to a 2$\times(4{+}n{+}k)$ matrix
\begin{align}
\mathbb{C}^{2\times (4|4)}\ni  \quad 
 \left(
 x| \theta
 \right)\ \rightarrow \ X=\left(
x, \xi
\right)=\left(
x, \theta  \phi
\right)\quad \in \mathbb{C}^{2\times(4{+}n{+}k)}
\label{bos-X}
\end{align}
where $x$ is the $2\times4$ matrix representing Minkowski space,
$\theta$ the $2\times4$ fermionic matrix (the fermionic part of
super Minkowski space)  and $\phi$ is a supplementary Grassmann odd
$4\times (4{+}n{+}k)$ matrix, which will be independent of the
space-time point. Thus  $X$ is a Grassmann even (bosonic)
$2\times(4{+}n{+}k)$-dimensional matrix.  Furthermore this matrix $X$ has a local
$Gl(2)$ acting  on the left, inherited from that of the supermatrix,
and thus has the natural 
interpretation of a two-plane in $4{+}n{+}k$ dimensions. We call this bosonised
super-Minkowski space $\M_b:=\Gr(2,k{+}n{+}4)$.
The correlahedron space 
itself is a subset of $\Gr(k{+}n,k{+}n{+}4)$, the space of $k{+}n$-planes, $Y$, in
$k{+}n{+}4$ dimensions.

We will use the following indices on the bosonised twistor space, space-time and -hedron space
\begin{align}
  Z_{i'}{}^{\cA'} &= (z_{i'}{}^A,\zeta_{i'}{}^{p'}) \, ,\quad &    X_{i\alpha}^\cA &= (x_{i\alpha}{}^A,\xi_{i\alpha}{}^p)\, ,\notag\\
  i'&=1\dots n',\;\;  p'=1 \dots k',\quad \cA'=(A,p'), & i&=1\dots n,\;\; p=1 \dots n{+}k ,\quad \cA=(A,p),\notag\\
  Y_{p'}^{\cA'}&\in Gr(k',4+k') &  Y_p^\cA&\in Gr(k+n,4+k+n)\, ,
\nonumber \\
 & \qquad \qquad\qquad \qquad A=1\dots 4,   \qquad \alpha=1,2\,.  \label{bos-table}
 \end{align}
Here the primed indices will correspond to the amplituhedron case and the unprimed to the correlahedron.  The index $A$ is for the  bosonic twistor coordinates, and $\alpha$ for homogeneous coordinates $\sigma^\alpha$ on the line in twistor space corresponding to the point $X$.  In certain $GL(2)$ gauge fixings it can be identified with a two-component self-dual spinor index.

Symmetries: for the amplituhedron we have a local $GL(k')$ acting from the left on the $p'$  index (corresponding to a different choice of basis
for the $k'$-plane $Y$ in $Gr(k',k'+4)$) and $n$ $GL(1)$s acting
scaling each $Z_i$.  We also have  a global $GL(k'+4)$ acting simultaneously on the right on the $\cA'$ index that $Y,Z$ carry (and the $k'+4$ space). For the correlahedron we analogously have 
 a local $GL(n+k)$ acting from the left on $Y$, and a global $GL(4+n+k)$ acting simultaneously on the right of $Y$ and $X$. In addition there is also a local $GL(2)^n$ with each $GL(2)$ acting on the $\alpha$ index of $X_{i\alpha}$ and corresponding to simply changing the choice of basis for each of the 2-planes $X_i$. 

\subsection{Bosonised correlator potentials}
\label{potentials}

We will consider 
correlators $\la \cO_1 \ldots \cO_n\rangle$ where $\cO_i$ is the
super-BPS operator whose leading part is $\tr
((y_i\cdot\Phi(x_i))^2)$.  Here the $y_i$ are skew matrices over the
four component $R$-symmetry indices that have rank two  and the $x_i$
are points in Minkowski space.  The supersymmetric extension extends
this to a function on analytic superspace~\cite{Galperin:1984av,Howe:1995md},
however in~\cite{Chicherin:2014uca} an alternative formulation for the
supersymmetric extension of the chiral correlator was found (see~\cite{Chicherin:2016fac,Chicherin:2016fbj,Chicherin:2016soh} for
the extension to the full non-chiral case). This describes
the correlator in terms of a potential $\cG_{n}$, a function of
$(x,\theta)$ in chiral super Minkowski space related by
$$\la \cO_1 \ldots \cO_n\rangle
=\left(\prod_{i=1}^nD_i^4\right)\cG_{n}(x_i,\theta_i)
$$
where 
$$
D_i^4:=y_i^{IJ}y_i^{KL}\p_{\theta^{\alpha I}_i}\p_{\theta^{\beta J}_i}\p_{\theta^{K}_{i\alpha}}\p_{\theta^{L}_{i\beta}}\, .
$$
 The correlator decomposes into irreducible parts  of degree $4k$ in
 the $\theta$s, and the corresponding potentials we denote $\cG_{n;k}$
 (which thus have degree $\theta^{4(n+k)}$).

We bosonise the dependence on the $\theta$s as in
(\ref{bos-X},\ref{bos-table}) to lift  $\cG_{n;k}$ to a function
$G_{n;k}(X_1,\ldots , X_n)$ 
defined on $n$ copies of $\M_b=Gr(2,4{+}n{+}k)$ where the $k$ corresponds to the fermionic degree $4(n+k)$.  

Now the potential  $\cG_{n;k}$ need not be gauge invariant, although the
correlator will be after the differentiation. 
In the twistor Feynman diagram formalism arising from the twistor
action, this potential is interpreted as a correlator of certain `log
det d-bar' operators based on lines in twistor space.  These operators
are not gauge invariant, although become so when differentiated by the
$D^4_i$ when they become  
the gauge invariant super-BPS operators $\cO_i$.
In the diagram formulation based on an axial gauge, the gauge
dependence will manifest itself in dependence on the reference twistor
$Z_*$.  We will suppress this differentiation in the following and
indeed there appears to be a natural $Z_*$ independent
volume form in correlahedron space, $G_{n;k}$ obtained directly from
analytic superspace expressions.

In the limit where $n'$ of the $x_i$ lie on a lightlike polygon, when
multiplied by $\prod_{i=1}^{n'} (x_{i\, i+1}^2)$ this correlator
degenerates into the loop integrand for the supersymmetric light-like
Wilson-loop at loop order $n-n'$.  Via the amplitude-Wilson-loop
duality this provides the afore-mentioned link to amplitudes in the planar limit. However, the Wilson-loop to which it degenerates is in the adjoint rather than fundamental representation and so gives the square of that in the fundamental that corresponds to the amplitude.  
We will see that the correlahedron  degenerates geometrically to give the squared amplituhedron in this limit.

\subsection{Correlahedron and amplituhedron   forms}
The correlahedron lives in the Grassmannian $\Gr(n{+}k,4{+}n{+}k)$, a
$4(n{+}k)$ dimensional space whose  points are represented by the $(4{+}n{+}k)\times (n{+}k)$  matrix $Y_p^\cA$ defined up to $GL(n+k)$ acting on the $p$-index.

The potential $G_{n;k}$ is given by  a volume form $\Omega_{n;k}(Y,X_i)$ on this space.  This gives rise to $G_{n;k}$ by the formula
\begin{equation}
G_{n;k}(X_i):= 
\int \Omega_{n;k}(Y,X_i)\delta^{4(n+k)}(Y;Y_0)
\end{equation} 
where  $\Omega_{n;k}(Y,X_i)$ is a $4(n+k)$- form on $\Gr(n{+}k,4{+}n{+}k)$ and 
\begin{equation}
\delta^{4(n+k)}(Y;Y_0):=\int d^{(k+n)^2} \rho^r_s \det (\rho)^4 \delta (Y_r-\rho_r^s Y_{s0})\, ,\quad \mbox{and} \quad  Y_0=\begin{pmatrix}
0_{4\times(n+k)}\\ 1_{(n+k)\times (n+k)}
\end{pmatrix}\, .
\label{delta-Y-Y0}
\end{equation}
In this formula, $\rho$ is a dummy variable that picks out the additional $(n+k) \times (n+k)$-components of $Y$ and takes their determinant which will then provide the bosonised form of the fermionic delta functions.

Similarly the (square of the) amplituhedron lives in the Grassmannian $\Gr(k,4{+}k)$ at tree-level with analogous formulae to the above with  
\begin{equation}
A_{n;k}(Z_1,\ldots,Z_n)=
\int \Omega_{n;k}(Y,Z_i)\delta^{4)}(Y;Y_0)\, ,
\end{equation}
with an analogous description for the loop integrand that we shall detail later.

Thus the key information of the correlator/amplitude is encoded in the volume
form $\Omega$. We first remark that there is a natural weighted volume form on $\Gr(k,4{+}k)$ of weight $k(4{+}k)$ that can be written as 
$$
\langle Yd^4Y_1\rangle \ldots \langle Y d^4 Y_k\rangle
$$
and similarly on $\Gr(n{+}k,4{+}n{+}k)$.    However, the overall expression must have weight zero in both $Y$ and the $X_i$ and $Z_i$.  The remaining factor that must balance the weights is proposed to  be characterized by
its poles, although  we will first find a representation as a sum of
Feynman diagrams, albeit in a gauge dependent form when it comes to
the correlator potential. This remaining factor essentially {\em is}
the bosonised correlator after putting $Y\rightarrow Y_0$ (which can
be done using the global $GL(n{+}k{+}4)$ symmetry). So for the amplitude
\begin{align}
 \Omega_{n;k}(Y,Z_i) = \langle Yd^4Y_1\rangle \ldots \langle Y d^4
  Y_k\rangle \times A_{n;k}(Y,Z_i), \qquad \qquad
  A_{n;k}(Z_i)=A_{n;k}(Y_0,Z_i) 
\end{align}
and for the correlator
\begin{align}
 \Omega_{n;k}(Y,X_i) = \langle Yd^4Y_1\rangle \ldots \langle Y d^4
  Y_{n+k}\rangle \times G_{n;k}(Y,X_i), \qquad \qquad
  G_{n;k}(X_i)=G_{n;k}(Y_0,X_i) \ .
\end{align}

\section{Hedron  geometry}

\label{sec:hedron-geometry}

In the previous section we saw that correlators, amplitudes
(possibly squared) and their loop integrands can be encoded in terms
of volume forms on respectively $\Gr(n{+}k,4{+}n{+}k)$ and
$\Gr(k,4{+}k)$.  A key aspect of the amplituhedron programme is
that these forms should be uniquely determined by the `hedron' geometry.

In this section we first review  the main features of the
amplituhedron as a geometrical object following~\cite{Arkani-Hamed:2013jha}. We then introduce a larger object in the same
space, $\Gr(k,4{+}k)$ which we call the ``squared amplituhedron'' and
which was been hinted at in~\cite{Arkani-Hamed:2017vfh}. This corresponds to the
square of the amplitude.
Finally we propose a new geometric object, the correlahedron, a subspace
of the higher dimensional Grassmannian $\Gr(n{+}k,4{+}n{+}k)$, and
which should correspond to the correlator.

\subsection{Amplituhedron}
\label{sec:amplituhedron}

The first definition of the amplituhedron is as the image of the positive
Grassmannian $Gr^+(k,n)$ of \emph{positive} $k$-planes in $n$ dimensions, into $\Gr^+(k,4{+}k)$.  Positive here means that all ordered $k\times k$ minors
are non-negative. The map from $Gr^+(k,n)$ to $Gr^+(k,k+4)$ follows from a linear map from $n$ to $k{+}4$
dimensions given by the external
kinematic data in the form of the $n$ bosonised momentum twistors  $Z^{\cA'}_i$ an $n\times (k{+}4)$ matrix.  The matrix $Z_{i}{}^{\cA}$ also has to be positive, ie
all its ordered maximal minors must be positive.
 In summary, the amplituhedron is the set
\begin{align}
  \label{eq:35}
\text{amplituhedron}_{n;k}(Z)=  \Big\{ Y \subset \Gr(k,4{+}k): \ Y^{\cA'}_{p'}=
  C_{p'}^iZ_i^{\cA'} \text{ for }  C \in 
  Gr^+(k,n) \Big \}\ .
\end{align}
One way to give an explicit description of this positive geometry is
via a BCFW decomposition of the amplitude in the Grassmannian \cite{ Mason:2009sa,Bourjaily:2013mma}.  
It is proposed that this geometric image uniquely determines the
volume form $\Omega$ as the unique holomorphic volume form of
$\Gr(k,4{+}k)$ that has logarithmic singularities on the boundary of
the region (and no singularities inside).

The above is the tree-level amplituhedron. At $\ell$-loops
there is an analogous object in which the Grassmannian $\Gr(k,4{+}k)$  
is supplemented by $ \ell$ 2-planes orthogonal to
$Y$.
The superamplitude  is then given as the 
differential form $\Omega$ that has logarithmic divergences on the boundary of this
amplituhedron. 
For more details of the amplituhedron see~\cite{Arkani-Hamed:2013jha}.

The 
above  definition is  somewhat implicit.   In general the map from $C$
to $Y$,  $Y=CZ$ is a projection  from a higher dimension,
that  maps many points to the same point. It is
difficult to extract an explicit logarithmic form (and hence the
amplitude) directly from the geometry without the original BCFW decomposition in the Grassmannian.
The definition~\eqref{eq:35},  together with the positivity of the
external data,  implies however the explicit 
$\Gr(k,4{+}k)$ constraints 
$$
\langle Y Z_{i{-}1}Z_iZ_{j{-}1}Z_j \rangle
>0\, ,
$$
where here $\langle \ldots\rangle$ is the skew form over $\R^{4+k}$ with $4+k$ arguments and 
$$
\langle Y ABCD\rangle:=\langle Y_1 \ldots Y_k\, ABCD\rangle\, .
$$
These 
constraints do indeed encode the location of the physical singularities but are not sufficient to fully specify the 
amplituhedron and in~\cite{Arkani-Hamed:2017vfh} a
further topological condition is understood to be required in addition.

\subsection{Squared amplituhedron}
\label{sec:squar-ampl}

The above discussion leads us to consider the subspace of $\Gr(k,4{+}k)$
defined simply by the  inequalities:
\begin{align}
  \label{eq:36}
 \text{squared amplituhedron}_{n;k}(Z)=  \Big\{ Y \in \Gr(k,4{+}k): \
  \langle Y Z_{i-1}Z_i Z_{j-1}Z_j \rangle > 0  \Big \}\ .
\end{align}
We call this the {\em squared amplituhedron} on the basis of the conjecture that this indeed gives the square of the amplitude. It lies in
the same space, $\Gr(k,4{+}k)$, as the amplituhedron and indeed contains
the amplituhedron, but  it is defined
by explicit constraints  in $\Gr(k,4{+}k)$ (without the additional
topological condition specifying the amplituhedron itself).
 This
explicit definition makes the squared amplituhedron much easier to use
in practice.

 Indeed we 
find in a number of examples that  
the logarithmic volume form associated with this region gives the
square of 
the (bosonised) superamplitude. The square of the superamplitude  of
Grassmann degree $4k$ is:
\begin{align}
  (A^2)_{n;k} = \sum_{k'=0}^k A_{n;k'}A_{n;k{-}k'}\ ,\label{eq:40}
  \end{align}
(obtained simply by expanding the square as a sum over $k'$ and taking
the relevant piece).
In section~\ref{sec:computations} we give a concrete
practical method (for small $n,\ell$) for obtaining the differential
form, and hence the superamplitude, from the squared amplituhedron using a
cylindrical decomposition.

The squared amplituhedron also extends to loop level. The $\ell$-loop
squared 
amplituhedron  is a subspace of the space of $k$-planes $Y \in \Gr(k,k{+}4)$ together with
$\ell$ complementary $2$-planes in $\R^{4+k}$, $\cL_i\in \Gr(2,4+k),
i=1,..,\ell$, subject to the following constraints
\begin{align}
  \label{eq:38}
& \text{squared amplituhedron}_{n;k}^{(\ell)}(Z)\notag\\
=&
  \Big\{ (Y,\cL_1,..,\cL_\ell) : \
  \langle Y Z_{i-1}Z_i Z_{j-1}Z_j \rangle > 0,\  \langle Y Z_{i-1}Z_i
  \cL_j \rangle > 0,\  \langle Y \cL_i \cL_j \rangle > 0  \Big \}\ .
\end{align}
The logarithmic differential form on this region gives the square of
the 
superamplitude at Grassmann degree $k$ and perturbative order $\ell$,
explicitly  it gives the combination:
\begin{align}
  \label{eq:39}
  (A^2)^{(\ell)}_{n;k} = \sum_{\ell'=0}^\ell \sum_{k'=0}^k A_{n;k'}^{(\ell')}
A_{n;k-k'}^{(\ell-\ell')}\ .
\end{align}
In section~\ref{sec:computations} we illustrate this squared
amplituhedron in some highly non-trivial
examples.

\subsection{Correlahedron}
\label{sec:correlahedron}

More importantly for this paper, the squared amplituhedron lends
itself to a natural generalisation, the {\em
  correlahedron}, on the basis of the conjecture that it should  yield the stress-tensor
correlator. We propose the correlahedron as a geometrical object lying
inside the 
space of $(k{+}n)$-planes in $\R^{4+n+k}$,  
$\Gr(n{+}k,4+n{+}k)$,  specified by the inequalities
\begin{align}\label{eq:41}
 \Big\{ Y \in \Gr(n{+}k,n{+}k{+}4): \
  \langle Y X_i X_j \rangle > 0  \Big \}\ .
\end{align}
Here the external data $X_i,
i=1,..,n$ are themselves 2-planes, $X_i \in \Gr(2,n{+}k{+}4)$, and are
equivalent to points in chiral superspace. 

It is the purpose of the rest of this paper to motivate
and give evidence for the correlahedron. 
We will begin in the next section by motivating the choice of space in which the
correlahedron lives, $\Gr(n{+}k,n{+}k{+}4)$, from an
algebraic point of view, starting with the formulation of
correlators using Feynman rules in twistor space~\cite{Chicherin:2014uca}.

\section{Hedron volume forms}

\label{sec:hedron-expressions}

We now describe the correlahedron volume forms (bosonised correlators) in
$\Gr(n{+}k,n{+}k{+}4)$ from a purely algebraic and analytic  perspective,
translating expressions found both from analytic superspace bootstrap techniques as well
as from  twistor space Feynman rules into the correlahedron space we
propose.  For
the correlator the expressions arising from twistor Feynman rules will
not be gauge invariant, however those arising from analytic superspace
bootstrap expressions  are, and this shows that there is nevertheless
a unique 
expression in $\Gr(n{+}k,n{+}k{+}4)$ which we propose
to be uniquely defined by the correlahedron geometry described in the
previous section.

\subsection{Hedron expressions from twistor space Feynman diagrams}

Here we explain how the amplituhedra and correlahedra  Grassmannians  described above arise
from considering Feynman diagrams in twistor space.  The key result is
that there will be a volume form $\Omega_\Gamma$ for each Feynman
diagram $\Gamma$ in twistor space.  The $\delta^{4(n+k)}(Y;Y_0)$ will
be seen to arise automatically from the product of propagators in a
diagram.  Each propagator will provide one physical singularity, but
there will be plenty of spurious singularities in each diagram, that
must cancel in the sum for the final correlator or Wilson loop. 

The twistor space Feynman rules are described for holomorphic Wilson loops in \cite{Mason:2010yk,Adamo:2011dq,Lipstein:2012vs} and the most developed version for the correlators can be found in \cite{Chicherin:2014uca}.
In this context we will use the amplitude/Wilson-loop duality to give
amplituhedron and squared amplituhedron expressions.   This is equivalent to using a momentum twistor formulation of the amplitudes. Furthermore, the polygonal lightlike Wilson-loop in space-time or region-momentum space will be understood as a holomorphic Wilson-loop for a polygonal loop in momentum twistor space.

 The diagrams contributing to the $\ell$-loop integrand of a holomorphic Wilson loop in twistor space depends on $n'$ twistors $Z_{i'}$ forming the vertices of the polygon in twistor space that corresponds to the edges of the light-like polygonal Wilson-loop in space-time, together with $\ell$ lines in twistor space corresponding to points in region momentum space for the loop integrand\footnote{When the loop integrand is obtained in this way, the region loop momenta come with fermionic coordinates also that need to be integrated out as part of the loop integration, but are part of the supersymmetric correlator.}.  
 We will take all our diagrams to be planar (firstly in order that the amplitude/Wilson-loop duality should hold, and to avoid more complicated rules associated with the colour structure).  At N$^{k'}$MHV degree there should be $2\ell + k'$ propagators connecting the lines and polygon.  Correlators are computed using essentially the same rules except that the propagators simply connect a collection of $n$ lines together.  In this case, it is said to have MHV degree $k$ when there are $n+k$ propagators as each line must have at least two propagators ending on it.
  In the light-like limit, $n'$ of these $n$ lines will  form the sides of the polygon and $n-n'=\ell$ the loop integrand points.  In this limit, the diagrams correpond to the amplitude$^2$ when the planar representation of the diagram extends both outside and inside the Wilson-loop, but reduces to the amplitude itself when only diagrams inside the polygon are allowed. 

In the following, the simplest case treated first is that for the correlator, where only lines in twistor space are needed connected by propagators.  The log-det operator insertions give rise to `MHV vertices' on these lines with a Parke-Taylor structure.  We can then incorporate a holomorphic Wilson-loop in twistor space essentially by regarding the edges of the Wilson loop to carry MHV vertices connected together without propagators around the polygon.

\subsubsection{Super twistor space Feynman rules}

Points in chiral superspace correspond to lines in $\CP^{3|4}$ spanned by the pair of twistors $X_{i\alpha}^\cA$, $\alpha=1,2$ where points on the line are parametrized homogeneously by $\sigma^\alpha$ by $Z_i(\sigma)=\sigma^\alpha X_{i\alpha}^\cA$.    
When we reduce to a  Wilson loop, we take the lines $X_{i'}$ $i'=1,\ldots, n$ to intersect in a polygon, but then we must integrate out $4n'$ superfluous fermionic coordinates (the $n'$ lines have $8n'$ fermionic coordinates, whereas the $n'$ twistors only $4n'$, so we require the identification of the fermionic parts of $Z_i$ as a point on $X_i$ with those on the point $X_{i+1}$.

The propagator connecting twistors $Z$ and $Z'$  corresponds to the delta-function 
\begin{equation}
\Delta(Z,Z') :=\int\frac{1}{\mathrm{vol} \, Gl(1)}\frac{dr}{r} \frac{ds}{s}\frac{dt}{t} \delta^{4|4}(rZ_*+sZ + tZ' )\, .
\label{propagator}
\end{equation}
To divide by vol $Gl(1)$ we can simply set one of the parameters $r,s,t$ equal to some constant, but it will be convenient to keep the scalings in play.  
In a diagram a propagator will connect a line
$X_i$ at the point $Z_i(\sigma_{ij})$ to another $X_j$ at the point $Z_j(\sigma_{ji})$.  
Each line $X_i$ supports a vertex corresponding to a `log det d-bar' operator on the line in twistor space that can in practice be thought of as an MHV vertex with as many legs as propagator insertions on the line.  If the number of propagator insertions is $M_i$, then the insertion points are given by $Z_i(\sigma_m)$, $m=1,\ldots ,M_i$ cyclically ordered by the planarity of the diagram.  The vertex requires an integration  over the insertion points $\sigma_r$
\begin{equation}
 \int \frac{1}{M_i \text{Vol}(GL(2)\times GL(1)^{M_i})}\prod_{m=1}^{M_i}\frac{d^2 \sigma_m}{(\sigma_m,\sigma_{m+1})} ,\label{dlog-vertex}
\end{equation}
where the integration points are understood projectively, hence the $GL(1)^{M_i}$ and the $GL(2)$ acts on the $\sigma_r$ and the $\alpha$ index on $X$.  The $GL(2)$s can all be fixed by setting $X_{i\beta}^\cA=(\delta_\beta{}^\alpha,x^{\dot\alpha}_{i\beta},\theta_{i\beta}^I)$ although in the Wilson loop context other gauge fixings can be more helpful. The $GL(1)$s in \eqref{dlog-vertex}  reflect the fact that the $\sigma$ integrals are projective.  However, the parameters $s$ and $t$ in \eqref{propagator} provide scalings for the $\sigma$s, otherwise said the $GL(1)$ quotients in \eqref{dlog-vertex} can be used to fix the $s$ and $t$ parameter  integrals in the propagators so that the $s$ in $s Z_i(\sigma_{ij})$ defines the scale of  $\sigma_{ij}$.  There is precisely one such $GL(1)$ for each of the two insertions of each propagator in the vertex  and so we set $s=t=1$.  The remaining $GL(1)$ in the propagator definition can be used to fix $r$ to be constant.  It nevertheless has nontrivial weight so we will not set it equal to one, but keep it in the formulae.  

With this gauge fixing, the propagator becomes
$\bar\delta^{4|4}(rZ_*+\sigma_{ij}\cdot X_i + \sigma_{ji}\cdot X_j)$
where the scaling integrals are now absorbed  into those for the $\sigma$s at each vertex and $r$ is an arbitrary nonzero constant that will not affect the final answer.
In the case of a correlator, the diagram's contribution $\cG_\Gamma$ to the potential  $\cG$ for the correlator is
\begin{equation}
\cG_\Gamma =  \int \prod_{i=1}^n \prod_{m_i=1}^{M_i}\frac{d^2 \sigma_{m_i}}{(\sigma_{m_i},\sigma_{m_i+1})}  \prod_{p=1}^{n+k}  \bar\delta^{4|4}(r_p Z_*+\sigma_{i_pj_p}\cdot X_{i_p} + \sigma_{j_pi_p}\cdot X_{j_p})\, .
\label{susy-diagram}
\end{equation}
Here the $\sigma$s are indexed in two ways, firstly by their locations $m_i$  on the $i$th vertex and secondly at the ends $i_p$ and $j_p$ of the $p$th propagator.\footnote{The integrations over the $\sigma$s can then all be done explicitly against the delta functions  with solution 
$$ 
\sigma_{ij\alpha}=\frac{(X_{i\alpha} Z_* X_{j1}X_{j2})}{X_i\cdot X_j}\, .
$$
Here $(Z_1 Z_2 Z_3 Z_4)$ is the skew form on the bosonic parts of the four twistors.  In doing these integrations against the delta functions, we obtain a Jacobian factor of $X_i\cdot X_j$ in the denominator for each propagator. }

It was shown in \cite{Chicherin:2014uca} that once the formulae have been differentiated by the product of the $D_i^4$, diagrams with two adjacent  propagators connecting $X_i$ to $X_j$ automatically vanish so we lose nothing by ruling out such diagrams ab initio.  It was further shown that spurious singularities associated with the $(\sigma_r \sigma_{r+1})$ factors in the Parke-Taylor denominators cancel via a process of three-way cancellation.  This latter property is no longer guaranteed without the $D^4_i$ differentiations.

We now discuss the extension of the Feynman diagrams to the holomorphic Wilson loop and hence amplitude (perhaps squared).  When $n'$ lines $X_{i'}$, $i'=1\ldots n'$ intersect so that $X_{i'}\cap X_{i'-1}=Z_i$ we obtain a polygon in twistor space with vertices $Z_{i'}$.  It was shown in \cite{Adamo:2011dq} that as this limit is approached, when multiplied by $\prod_{i'=1}^n X_{i'-1}\cdot X_{i'}$,  the Feynman  diagrams become those for the adjoint holomorphic Wilson loop in twistor space which is the same as the adjoint super Wilson-loop in chiral super Minkowski space, that can in turn be identified with the square of the amplitude.

In more detail, the Feynman diagrams for the adjoint holomorphic Wilson loop now has two types of vertices, the lines  $X_{i'}$ that form part of the polygon, and those that do not (these latter in this context correspond to the loop variables in the amplitude interpretation or Lagrangian/stress-energy insertions in the Born approximation).  
In taking the lightlike limit, we simply omit the $n'$ propagators that connect the now joined consecutive $X_{i'}$ and $X_{i'-1}$.  
However,  we do keep the vertices at the $X_{i'}$ including the connections between the $X_{i'}$ in the Parke-Taylor factors.  These can be gauge fixed using the $GL(2)$ in \eqref{dlog-vertex} for the sides of the polygon so that $\sigma\cdot X_{i'}=\sigma^0 Z_{i'}+ \sigma^1 Z_{i'+1}$ and the $\sigma$ at $Z_{i'}$ is $\sigma_0=(1,0)$ and that at $Z_{i'+1}$ is $\sigma_{I_{i'}+1}=(0,1)$ when there are $I_{i'}$ propagators attached to the $i'$ edge of the Wilson loop inside the polygon and $O_{i'}$ outside.   Thus it gives rise to a factor
\begin{equation}
 \int \prod_{m=0}^{I_{i'}+O_{i'}+1}\frac{d^2 \sigma_m}{(\sigma_m,\sigma_{m+1})} \, . \label{WL-edge}
\end{equation}
where we have taken the $GL(1)$ scalings to be fixed against the
propagators as above (although note that this is a different gauge
fixing for the Feynman diagrams for the Wilson loop to those given in
\cite{Adamo:2011dq,Lipstein:2012vs} say). The distinction between the diagrams considered here is that here the planar diagrams have propagators and vertices both outside and inside the Wilson loop, whereas for the Wilson-loop in the fundamental representation and hence the amplitude, they are purely inside.

In order to obtain the the loop integrand itself, we must eventually integrate out all the fermionic $\theta$ variables at the $X_i$ when $X_i$ is a region loop variable (if we only do $D_i^4$ we are essentially obtaining the Born level correlator of a Wilson loop with $\tr \Phi^2$ rather than a Lagrangian insertion corresponding to a loop integrand point).

\subsubsection{Bosonisation of Feynman diagrams in the correlahedron space}
The Hodges bosonisation of the fermionic variables yields 
$$
\delta^{0|4}( \chi^I)=\int ( \chi \cdot \phi)^4 d^4 \phi\, ,
$$
and this motivates the introduction of new such variables $\phi_I{}^p$, four for  each propagator
$p=1,..,n{+}k$.  This gives the 
new bosonic variables $\zeta^p=\chi\cdot\phi^{p}$ In the amplitude formula above we will then replace the $a$th delta function by 
$$
\delta^{4|4}( \cZ) \rightarrow  (\zeta^p)^4\prod_{A=1}^4 \delta(Z^A)
$$
  We retrieve the original $\delta^{4|4}(\cZ)$ by substituting in  and integrating out the $\phi_{I}{}^p$. 
With this \eqref{susy-diagram} becomes
\begin{equation}
G_\Gamma =  \int \prod_{i=1}^n \prod_{m_i=1}^{M_i}\frac{d^2 \sigma_{m_i}}{M_i(\sigma_{m_i},\sigma_{m_i+1})}  \prod_{p=1}^{n+k}  (y^p)^4 \delta^{4}(r_pZ_*+\sigma_{i_pj_p}\cdot X_{i_p} + \sigma_{j_pi_p}\cdot X_{j_p})\, .
\label{bos-diagram}
\end{equation}
where $y^p=\sigma_{i_pj_p}\cdot \xi_{i_p} + \sigma_{j_pi_p}\cdot \xi_{j_p}$.

We can now define the map from the $\sigma$ parameters to the correlahedron Grassmannian by
\begin{equation}
Y_p^\cB=
r_p Z_{*}^\cB+ \sigma_{i_pj_p}\cdot X_{i_p}^\cB + \sigma_{j_pi_p}\cdot X_{j_p}^\cB
\, . \label{embed}
\end{equation}
where here now $\cB=(B,q)=(1\ldots 4, 1\ldots n+k)$ (we could here include a $\zeta_*^p$ part of $Z_*$ in the same way as we could have had a fermionic part of the original reference twistor).

With this, the product over $p$ on the right hand side of \eqref{bos-diagram} becomes $\prod_p (y_p^p)^4 \delta^4(Y^A_p)$.  However, we obtain the same formula for the super-amplitude if we replace this expression by $\delta^{4(n+k)}(Y;Y_0)$ as defined in \eqref{delta-Y-Y0}.    On performing the $\rho$-integral in \eqref{delta-Y-Y0}, this is equivalent to replacing   $\prod_p (y_p^p)^4$ by  $(\det\{y_p^q\})^4$.  This will yield the same super-amplitude up to some numerical factor  because,  after inserting $y_p^q= \eta^q \phi_p$, the transform back to the supersymmetric correlator/amplitude picks out the coefficient of the top power of  
$\phi$s and this will of necessity be the top power of the $\eta$s that provide the arguments of the desired $\delta^{0|4}$s in \eqref{susy-diagram}. Thus the only required check is that the numerical factor is not zero, which can be done by hand.

Thus, identifying the $\delta$-functions arising from the propagators with  
$\delta^{4(n+k)}(Y;Y_0)$, we obtain the diagram's contribution to the correlator by the formula
\begin{equation}
G_\Gamma(X_i,Z_*):= 
\int \Omega_\Gamma(Y,X_i,Z_*)\delta^{4(n+k)}(Y;Y_0) \, .\label{diagram-corr}
\end{equation} 
where the  $4(n+k)$-form $\Omega_\Gamma$ on the correlahedron Grassmannian is the product of the vertices
\begin{equation}
\Omega_\Gamma (Y,X_i,Z_*):=\prod_{i=1}^n  \prod_{m=1}^{M_i} \frac{d^2\sigma_{ij_m}}{(\sigma_{ij_m}\sigma_{ij_{m+1}})} \, .
\label{diagram-form}
\end{equation}

This formula can be expressed in terms of the $Y$s by using \eqref{embed}.  We introduce the notation
$$
\langle Y ABCD\rangle:= (Y_1 Y_2 \ldots Y_{n+k}  ABCD)
$$
where on the right hand side $(\ldots )$ denotes the natural skew bracket over $4+n+k$ space of objects with an $\cA$-index.  
This expression taken with $ Y_p$ as one of $A,B,C,D$  must vanish, so 
$$
0=\langle Y Y_p X_{i1} X_{i_2} X_{j\alpha}\rangle=r_p\langle Y Z_*  X_{i1}X_{i2} X_{j\alpha}\rangle+ \sigma_{j_pi_p\alpha}\langle Y  X_i X_j\rangle\, ,
$$
yielding
\begin{equation}
\sigma_{j_pi_p\alpha}=-r_p\frac{\langle Y Z_*  X_{i_p1}X_{i_p2} X_{j_p\alpha}\rangle}{\langle Y  X_{i_p} X_{j_p}\rangle}\, .
\label{sigma}
\end{equation}
Similarly, taking the exterior derivative of \eqref{embed} (regarding $X_i$ and $Z_*$ as constants) and inserting the resulting equation 4 times into $\langle Y \ldots \rangle$ we find
$$
d^2\sigma_{j_pi_p}d^2\sigma_{i_pj_p}= \frac{ \langle Y d^4Y_p\rangle}{\langle Y  X_{i_p} X_{j_p}\rangle}\, .
$$

With this we can write the volume form as 
\begin{equation}
\Omega_\Gamma (Y,X_i,Z_*):=\prod_{i=1}^n  \prod_{m=1}^{M_i} \frac{1}{(\sigma_{ij_m}\sigma_{ij_{m+1}})} \prod_{p=1}^{n+k}\frac{ \langle Y d^4Y_p\rangle}{\langle Y  X_{i_p} X_{j_p}\rangle}\, .
\label{diagram-form-inv}
\end{equation}
In this we can see that we have one `physical' singularity for each propagator namely the $\langle Y X_i X_j\rangle$ and four spurious ones, essentially the adjacent
Parke-Taylor denominators (shared with the adjacent propagators at the vertex).\footnote{The denominator in \eqref{sigma} might be thought to affect the poles in $\langle YX_iX_j\rangle$ for each propagator, but these factors  cancel in the final formulae as there are fourth powers of the $\sigma$s in the $\det \{Y_p^q\}$ in the $\delta^{4(n+k)}(Y;Y_0)$ in \eqref{diagram-corr}.  This factor could for example be incorporated into $r_p$. }  We remark that the cancellation of these spurious singularities was identified in \cite{Chicherin:2014uca} as being between triples of diagrams that agree everywhere except on  a triangle between three vertices $X_i,X_j$ and $X_k$ with each diagram having two out of three propagators around the triangle.

\subsection{Invariant correlahedron expressions directly from
  correlators on analytic superspace}
\label{sec:expl-corr-form}

In the previous subection we translated Twistor Feynman expressions
directly into correlahedron form expressions. Unfortunately the
resulting expressions were not gauge invariant. However we find that
one can alternatively lift directly from the analytic superspace
expressions in a canonical gauge invariant way to obtain a unique
canonical correlahedron volume form 
for each correlator.

Many correlators have by now been constructed explicitly writing down
forms with the correct singularity structure and showing that they
satisfy appropriate consistency properties.  In particular, for
maximal $k=n-4$, they have been constructed up to $n=14$ (equivalent
to 10 loop four-point correlators), and the next to maximal  case,
$k=1$, $n=6$ has also been
constructed~\cite{GonzalezRey:1998tk,Eden:1998hh,Eden:1999kh,Eden:2000mv,Bianchi:2000hn,Eden:2011we,Eden:2012tu,Ambrosio:2013pba,Bourjaily:2015bpz,Chicherin:2015bza,Bourjaily:2016evz}.
We will see here that given these expressions, there is a simple
procedure to uplift them directly (and uniquely) into correlahedron
volume 
forms on $\Gr(n{+}k,4{+}n{+}k)$.  

We start with the simplest non-trivial correlator, the 5 point $k=1$
case $G_{5;1}$. The physical singularities are at $\langle Y X_i X_j\rangle$.   The correlahedron space for this correlator is $Y \in
\Gr(6,10)$.  There is essentially a unique  correlahedron form of weight zero in $Y$ and the $X_i$  with simple poles at the physical singularities.  It is given by
\begin{align}
  \label{eq:3}
\Omega_{5;1}(Y,X_i)=\frac{ \langle Y d^4Y_1\rangle \dots \langle Y d^4Y_6\rangle \langle X_1 X_2 X_3 X_4 X_5  \rangle^4}
{\langle{YX_1X_2}\rangle\langle{YX_1X_3}\rangle \dots \langle
  YX_4X_5\rangle}\ .
\end{align}

The next simplest cases to consider are the  maximally nilpotent
($k=n-4$) correlators. These   are
described in terms of 
a single function $f$~\cite{Eden:2000bk} which is a conformally covariant, permutation
symmetric function of $x_{ij}^2$
\begin{align}
  f^{(n{-}4)}(x_{ij}^2)\ .
\end{align} 
These functions are known explicitly for $n\leq 14$.
The corresponding correlahedron space for these correlators is $Y\in \Gr(2n-4,2n)$ and the correlahedron forms are
given simply as 
\begin{align}
  \label{eq:7}
 \Omega_{n;n-4}(Y,X_i)=  \langle Y d^4Y_1\rangle \dots \langle Y d^4Y_{2n{-}4}\rangle
  \times \langle X_1 \dots X_{n}  \rangle^4 \times f^{(n-4)}\Big( \langle Y X_i X_j \rangle\Big)\ .
\end{align}

So for example, for $n=5$, corresponding to the four-point one-loop
correlator, the function $f$ is
\begin{align}
  \label{eq:10}
  f^{(1)}(x_{ij}^2)=\frac 1{\prod_{1\leq i<j\leq 5} x_{ij}^2} 
\end{align}
and we see that making the replacement $x_{ij}^2\rightarrow \langle
YX_iX_j\rangle$,~\eqref{eq:7} correctly reproduces~\eqref{eq:3}.

The only non-maximally nilpotent correlator currently known explicitly is the
six point $k=1$ correlator. This was derived in analytic superspace
in~\cite{Chicherin:2015bza} and also has a straightforward lift to
a correlahedron 
volume form. This correlahedron lives in the space $Y \in \Gr(7,11)$ so all the
angle brackets are 11-brackets in the following expression.
Since we are considering 11-brackets but we have 12 points (6-space
time points) it is useful to label the 11-brackets by the missing
point. So we define
\begin{align}
  \label{eq:11}
  \langle \dots \rangle^{i \alpha} := \langle X_{11} X_{12} X_{21} 
  \dots \widehat{X_{i\alpha}} \dots X_{62}\rangle(-1)^{\alpha}\ .
\end{align}

The correlator $G_{6;1}$ was given  in~\cite{Chicherin:2015bza} in
terms of nilpotent superconformal invariants
$\cI^{ijkl;\alpha\beta\gamma\delta}$ in analytic superspace. Lifting
the correlator to the 11-dimensional correlahedron space, these
nilpotent invariants become the following product of 4 11-brackets: 
\begin{align}
  \cI^{ijkl;\alpha\beta\gamma\delta} = \langle \dots \rangle^{i\alpha} \langle \dots \rangle^{j\beta} \langle \dots \rangle^{k\gamma} \langle \dots \rangle^{l\delta} \ .\label{eq:37}
\end{align}

One interesting consequence of this correlahedron formulation of the correlator as an object in 11-dimensions is that it manifests highly non-trivial identities involving these invariants. It was observed in~\cite{Chicherin:2015bza} that the invariants satisfy the non-trivial identity
\begin{align}
  \sum_{i=1}^6  X_{i\alpha} \cI^{ijkl; \alpha\beta\gamma\delta}   = 0
\qquad (\text{for all } j,k,l,M,\beta,\gamma,\delta)\,,
\end{align}
which was found as a non-trivial consequence of superconformal Ward
identities. In the 11-dimensional correlahedron space this identity is
a straightforward consequence of a generalised Schouten identity in 11
dimensions 
\begin{align}
  \sum_{i=1}^6  X_{i\alpha} \langle \dots \rangle^{i\alpha}  =0\ .
\end{align}

The correlator itself as a correlahedron volume form  has the
representation 
\begin{align}\label{ABC}
   G_{6;1}^{(0)}=  \langle Y d^4Y_1\rangle \dots \langle Y d^4Y_7\rangle   \frac{A_2- 2 \, A_1 -8 \, B_2 }{\prod_{1 \leq i < j \leq 6} \langle Y X_iX_j\rangle}\,,
\end{align}
where we introduced the notation
\begin{align}\notag\label{AB}
{}&A_1=\langle Y X_{5\alpha} X_1 X_{6\gamma} \rangle   \langle Y X_{5\beta} X_2 X_{6\delta} \rangle \langle Y X_{3} X_5  \rangle\langle Y X_{4} X_{6} \rangle \cI^{5566;\alpha\beta\gamma\delta} + \text{$S_6$ permutations}\,,
\\[2mm]\notag
  {}&A_2=\langle Y X_{5\alpha} X_1 X_{6\gamma} \rangle   \langle Y X_{5\beta} X_2 X_{6\delta} \rangle \langle Y X_{3} X_4  \rangle\langle Y X_{5} X_{6} \rangle \cI^{5566;\alpha\beta\gamma\delta} + \text{$S_6$ permutations}\,,  \\[2mm]
  {}&B_2=\langle Y X_{4\alpha} X_3 X_{6\gamma} \rangle   \langle Y X_{5\beta} X_2 X_{6\delta} \rangle \langle Y X_{1} X_6  \rangle\langle Y X_{4} X_{5} \rangle \cI^{4566;\alpha\beta\gamma\delta} + \text{$S_6$ permutations}\,.
\end{align}
This form was directly lifted from the corresponding formula found
in~\cite{Chicherin:2015bza} in analytic superspace. 

It is clear from this example that the construction of
superconformal invariants on analytic superspace has a direct uplift
into the correlahedron space more generally. Indeed the invariants
$\cI^{ijkl;\alpha\beta\gamma\delta}$ are well-defined for $k=n-5$ for {\em any}
$n$ and have the natural uplift to correlahedron space given
by~(\ref{eq:37}).  But furthermore, the $k=1$ analytic superspace
superconformal invariants 
generalise to lower $k$. For
example, for 
$k=n-6$ the most general invariants on analytic superspace are of the
form
$\cI^{\{i_1i_2i_3i_4\}\{j_1j_2j_3j_4\};\{\alpha_1\alpha_2\alpha_3\alpha_4\}\{\beta_1\beta_2\beta_3\beta_4\}}$
which is symmetric under simultaneous interchange of $i_a \alpha_a$
with $i_b \alpha_b$ or separately $j_a \beta_a$
with $j_b \beta_b$.
These invariants have a natural uplift to correlahedron space:
\begin{align}
\cI^{\{i_1i_2i_3i_4\}\{j_1j_2j_3j_4\};\{\alpha_1\alpha_2\alpha_3\alpha_4\}\{\beta_1\beta_2\beta_3\beta_4\}} 
=\langle\dots\rangle^{i_1\alpha_1j_1\beta_1}
\langle\dots\rangle^{i_2\alpha_2j_2\beta_2}\langle\dots\rangle^{i_3\alpha_3j_3\beta_3}\langle\dots\rangle^{i_4\alpha_4j_4\beta_4}
  \ +\ \dots
\end{align}
where $\langle\dots\rangle^{i\alpha j\beta}$ is the $n{-}2$
bracket with $X_{i\alpha}$ and $X_{j\beta}$ missing and where we
sum over the 24 different possible simultaneous permutations of the $j_a\alpha_a$. 

Thus we see that, although the direct construction of the correlator
potential  in 
correlahedron space arising from the twistor Feynman rules outlined in
the previous section yields
an expression which explicitly depends on $Z_*$, nevertheless,
there is a canonical $Z^*$-independent uplift directly from an
analytic superspace 
expression to the correlahedron form. We conjecture that this
canonical form is uniquely defined by the correlahedron geometry as we
discuss further in section~\ref{sec:computations}.

\section{The lightlike limit on the correlahedron}
\label{sec:lightlike-limit}
As discussed previously, in the lightlike limit when consecutive space-time points become
lightlike separated,  the stress-tensor correlator reproduces the light like polygonal Wilson loop in the adjoint representation, and hence the
(square of the)
amplitude~\cite{Alday:2010zy,Eden:2010zz,Eden:2010ce,Eden:2011yp,Adamo:2011dq,Eden:2011ku}. One
can take all $n$ points lightlike separated around a polygon, in which case one gets the
corresponding tree-level amplitude. Or one can take a non-maximal
lightlike limit in which fewer than $n$ points are consecutively
lightlike separated. In this limit the resulting object is a square of a loop
level amplitude, with the remaining points corresponding to loop
variables. 

The reduction from the correlator to the amplitude squared is an explicit algebraic process.   We first take the limit as  $(x_i-x_{i+1})^2\rightarrow 0$ for $i=1, \ldots, n'$  (understood cyclically with $n'+1=1$) of $G_{n;k}(X_i) \prod_{i=1}^{n'} (x_i-x_{i+1})^2$ but we must then also integrate out half the fermionic dependence of the $X_{i'}$ $i'=1, \ldots n'$.  We wish to reduce the eight fermionic $\theta_{i'}$s for each $X_{i'}$ to the four fermionic variables for each twistor $Z_{i'}$ that make the corners of the corresponding polygon in twistor space.  When $n>n'$ we must furthermore remove all the fermionic dependence of the remaining $X_i$, $i>n'$ which then have the interpretation of region loop momenta for the loop integrand.

This apparently fairly complicated procedure, however has a very
simple and beautiful geometrical interpretation in correlahedron
space which we denote ``freeze and project''. Furthermore the same
geometric procedure acts both on the correlahedron {\em geometry} as well as
on the
corresponding {\em algebraic expressions}. This thus gives further
confirmation 
of  our
conjecture that the correlahedron  determines the correlator.

In this section we explain this lightlike limit 
on the correlahedron showing how it reproduces the
corresponding (squared) amplituhedron. We show this both geometrically as well
as algebraically.

\subsection{The maximal lightlike limit geometrically}
\label{sec:maxim-lightl-limit}

Taking the $n$-point lightlike limit of the correlator $G_{n;k}$ (we
will consider lower point lightlike limits shortly) has the following
geometric interpretation in correlahedron space. Recall that the 
correlahedron space is the subspace of the space of  $k{+}n$-planes, $Y$,
in a $4{+}k{+}n$-dimensional space bounded as~(\ref{eq:41})
\begin{align}
  Y \in \Gr(n{+}k,4{+}n{+}k): \qquad \langle Y X_i X_j \rangle >0 \quad
  \forall \ i,j=1..n\ .\label{eq:2}
\end{align}
The $n$-point lightlike limit is obtained by requiring $Y$ to simultaneously lie on
multiple boundaries $\langle YX_i X_{i+1} \rangle=0, i=1..n$ of the correlahedron.  
This can be done by freezing the first $n$ of the $Y_p$, i.e., $Y_i$, $i=1,\ldots n$ to lie respectively in the span of the consecutive $X_i \wedge X_{i+1}$.

To further reduce to the amplitude (squared) we need to reduce the
fermionic degree by $4n$ and hence the range of the $p$ index inside
$\cA$ to $p'=1\ldots k$.  
This  also has a natural geometric
interpretation for the 
correlahedron,  namely it corresponds to performing a projection
orthogonal to 
the $n$ frozen intersection points $Y_i$.
Thus the  $4{+}k{+}n$ dimensional space is projected  down to $4{+}k$
dimensions and  the $k{+}n$-plane in $4{+}k{+}n$ dimensions, $Y$,  is projected to a $k$-plane in $4{+}k$ dimensions. This $k$ plane gives the (square of the) amplituhedron.

In practical terms 
we can perform the freezing of  $Y$ to the boundary 
by choosing a $GL(n+k)$ basis so that  $Y =Y_1 \wedge ..\wedge Y_{n+k}$ with
\begin{align}\label{eq:1}
  Y_p &= \sigma_i^\alpha X_{i\alpha} - \tau_i^\alpha X_{i{+}1\,\alpha}
  \qquad \mbox{ for } p= i=1\dots n\ ,\notag \\
Y_{p}&=\hat Y_{p'} \qquad \qquad\qquad\qquad\quad  p=n+p', \quad p'=1\dots k\ 
\end{align}
for some parameters $\sigma_i^\alpha$, $\tau_i^\alpha$.
We then need to project onto the quotient by $Y_1, \dots
Y_n$.  In practice we can pick a basis  for the $k{+}n{+}4$ dimensional vector space
\begin{align}
\text{basis} = \Big\{Y_1, \dots, Y_n, e_1,\dots, e_{4+k}\Big\}\ ,
\end{align}
 where $e_1,\dots e_{4+k}$
are any $4+k$ vectors such that this yields an independent
basis.\footnote{Geometrically the span of the $e_i$ give the hyperplane on
which we are projecting. However the final result is independent of
this choice of hyperplane. }
We choose $\hat Y_{p'}$ to be a linear combination of the $e_{\cA'}$  in this
basis. 
The projection takes the form
\begin{align}
X_{i\alpha} \rightarrow \hat X_{i\alpha} \quad \text{where} \quad
 \hat X_{i\alpha}^\cA = \left\{
  \begin{array}{ll}
0 \quad &\cA =1,\dots,n\\
     X_{i\alpha}^\cA \quad &\cA=n{+}1,\dots, n{+}k{+}4
  \end{array}
\right.\ ,
\end{align}
in the basis $\{Y_1 \dots Y_n, e_1,\dots e_{4+k}\}$.

We can then define brackets in the obvious way on the hyperplane spanned by
$\{e_1,\dots,e_{4+k}\}$ and it is clear that
\begin{align}
  \langle \hat \cX  \rangle :=   \langle Y_1 \dots Y_n \cX  \rangle\ .
\end{align}
Here $\cX$ represents any collection of $4+k$ independent vectors, and
$\hat \cX$ the same vectors projected onto the hyperplane. 

After projecting out the $Y_i$, \eqref{eq:1} gives
$\sigma_i.\hat X_i =    \tau_i.\hat X_{i+1}$ so we can define 
\begin{align}
  Z_i:=\sigma_i.X_i &= \tau_i.X_{i+1} + Y_i  \ .\label{eq:28}
\end{align}
Then after the projection
\begin{align}
  \hat Z_i:=\sigma_i.\hat X_i = \tau_i.\hat X_{i+1}\ .
\end{align}
and  the projected planes $\hat X_i$  intersect each other
consecutively at $\hat Z_i$ in
the projected space.

Thus  freezing and projection yields a $k$-plane $\hat
Y$ living in the $4{+}k$ dimensional hyperplane spanned by  $\{ e_1,\dots
e_{4+k}\}$ and we have projected planes $\hat X_{i\alpha}$ in the same
$4{+}k$ dimensional space. Further we have   
\begin{align}
\langle Y X_i X_j\rangle= \left\{
  \begin{array}{ll}
    0&\quad |i-j|=1 \mod n\\ 
\frac{\langle \hat Y \hat Z_{i{-}1} \hat Z_i \hat Z_{j{-}1} \hat Z_j \rangle}
    {\tau_{i{-}1}.\sigma_i \,\tau_{j{-}1}.\sigma_j}&\quad
                                                          \text{otherwise}\ .
  \end{array}\right.\label{eq:30}
\end{align}
So the correlahedron space~\eqref{eq:2} reduces to%
\footnote{If we assume we have chosen $\sigma_i$ and $\tau_i$
  appropriately so that $\tau_{i-1}.\sigma_i>0$. Indeed different
  choices of signs here usually but not always yield  the same
  expressions for the correlator. In some cases one has to sum over
  different choices of signs (see section~\ref{sec:six-point-nnmhv}
  for example).  } 
\begin{align}
  \hat Y \in \Gr(k,4+k): \qquad \langle \hat Y Z_{i-1} Z_i Z_{j-1} Z_j
  \rangle >0
\end{align}
which is the squared amplituhedron~(\ref{eq:36}).

\subsection{Maximal lightlike limit on the hedron volume forms}\label{sec:lightl-limit-corr}

In the previous section we described the lightlike limit of the
correlahedron geometry as a freezing and projection of the space of
$Y$s.  Notice that this procedure is not singular as one might
expect. It is simply a restriction of the geometry to a partial
boundary, followed by a projection.

Here we give a simple algorithm for implementing this exact same
procedure 
directly on the expressions for the correlator as 
differential forms in correlahedron space. We find that this indeed
correctly reduces the correlator expressions to the correct amplitude
(squared) expressions.

The fully covariant correlahedron form should have simple poles at $\langle Y X_i X_{i+1}\rangle=0$ so we can write
\begin{align}\label{gnkdef}
\Omega(Y,X_i)=\frac{g_{n;k}(Y,X_i)}{\prod_{i=1}^n \langle Y X_i X_{i+1}\rangle} \prod_{p=1}^{n+k}  \langle Y d^4Y_p \rangle  \, ,
\end{align}
where $g_{n;k}$ has weight two in each $X_i$ and $- (n+k)(k+4)$ in $Y$.  When we freeze $Y$ as in~\eqref{eq:1} we find
\begin{align}
      \langle Y d^4 Y_i\rangle = \langle Y X_{i} X_{i+1} \rangle
      d^2\sigma_i d^2\tau_i \qquad i=1..n\ .
    \end{align}
Thus as $\langle Y X_{i} X_{i+1} \rangle \rightarrow 0$ in the
limit, it cancels the corresponding term in the
denominator of the correlahedron form to yield a finite result.
Inverting~\eqref{eq:28} we obtain
\begin{align}
  X_{i\alpha} = \frac{-\tau_{i{-}1\, \alpha}Z_i +\sigma_{i\,
  \alpha}(Z_{i-1}+Y_{i-1})}{\tau_{i-1}.\sigma_i}\ .
\end{align}
Projecting along $Y_i$ and correspondingly hatting the $Z$s in this expression sends%
\footnote{Note that this is not quite the same as replacing
  $X_{i\alpha} \rightarrow \hat X_{i\alpha}$ as the $Y_{i-1}$ remains
  on the right hand side and is not projected away. We need to do tis
  in order to make sense of the $k{+}n{+}4$-brackets.} 
\begin{align}
X_{i\alpha} \rightarrow \frac{-\tau_{i{-}1\, \alpha}\hat Z_i +\sigma_{i\,
  \alpha}(\hat Z_{i-1}+Y_{i-1})}{\tau_{i-1}.\sigma_i}\ .\label{eq:29}
\end{align}
 
The correlahedron form is then reduced to the amplituhedron squared form
by setting $Y$ to~\eqref{eq:1} and
$X_{i\alpha}$ to~\eqref{eq:29}  and finally leaving out the
$\sigma,\tau$ dependent factors:
\begin{align}\label{eq:42}
\frac{g_{n;k}(Y,X_i)}{\prod_{i=1}^n \langle Y X_i X_{i+1}\rangle}  \prod_{p=1}^{n+k}  \langle Y d^4Y_p \rangle & \rightarrow\notag\\
&\left(\prod_{i=1}^{n} {d^2 \sigma_i
  d^2\tau_i}\right) \left(\prod_{p'=1}^k\langle \hat Y d^4\hat Y_{p'}
  \rangle\right) g_{n;k}\Big(Y, \frac{-\tau_{i{-}1\, \alpha}\hat Z_i +\sigma_{i\,
  \alpha}(\hat Z_{i-1}+Y_{i-1})}{\tau_{i-1}.\sigma_i} \Big)\notag\\
&=\left(\prod_{i=1}^{n} \frac{d^2 \sigma_i
  d^2\tau_i}{(\tau_{i-1}.\sigma_i)^2} \right) \left(\prod_{i=1}^k\langle \hat Y d^4\hat Y_i
  \rangle\right) a_{n;k}(\hat Y,\hat Z_i)\notag\\
&\rightarrow \left(\prod_{p'=1}^k\langle \hat Y d^4\hat Y_i
  \rangle\right) a_{n;k}(\hat Y,\hat Z_i)
\end{align}
Here to go from the second to the third line, we used the fact that $g_{n;k}$ as defined in~\eqref{gnkdef} has homogeneity degree two in each $X_i$.   We have defined $a_{n;k}(\hat Y, \hat Z_i)= g_{n;k}(Y, -\tau_{i{-}1\, \alpha}\hat Z_i +\sigma_{i\,
  \alpha}(\hat Z_{i-1}+Y_{i-1}) )$ by this relation and this should
correspond to the square of the amplitude. In particular it should be
independent of $\sigma,\tau$: this is a direct consequence of the
amplitude/Wilson loop duality. 
This $\sigma,\tau$ dependence corresponds to choosing different points on
the boundary 4-planes to freeze $Y$.
 To go from the third to the fourth line
we simply drop the first factor that depends 
only on $\sigma,\tau$ which we are freezing.

\smallskip

As an explicit  example, take the expression for the correlahedron
form $G_{5;1}$~\eqref{eq:3} and perform the above freezing of $Y$ and
projection. We have  $Y=Y_1\wedge \dots \wedge Y_6 \ \in\ \Gr(6,10)$
and we freeze $Y_1,\dots,Y_5$ as in~\eqref{eq:1}, $Y_i =
\sigma_i^\alpha X_{i\alpha} - \tau_i^\alpha X_{i{+}1\,\alpha}$,
leaving $Y_6=\hat Y$ orthogonal. Then
  \begin{align}
 \prod_{i=1}^{6} \langle Yd^4Y_i \rangle    \frac{   \langle X_1 X_2 X_3 X_4 X_5 \rangle^4}{\langle YX_1X_2\rangle \dots \langle YX_4X_5\rangle}
    \xrightarrow[\text{project} X]{\text{freeze $Y$}} &\left(\prod_{i=1}^5
                                      {\frac{d^2\sigma_id^2\tau_i}{(\tau_{i-1}.\sigma_i)^2}}\right)
                                      \frac{  \langle Y d^4\hat
                                      Y\rangle \langle Y_1..Y_5 \hat Z_1
                                      .. \hat Z_5 \rangle^4}{\langle Y
                                      \hat Z_1\hat Z_2\hat Z_3\hat Z_4\rangle \dots
                                      \langle Y\hat Z_{5}\hat Z_1\hat Z_2\hat Z_3
                                      \rangle}\notag
    \\=&\left(\prod_{i=1}^5
         {\frac{d^2\sigma_id^2\tau_i}{(\tau_{i-1}.\sigma_i)^2}}\right)
         \frac{  \langle \hat Y d^4\hat Y\rangle \langle  \hat Z_1 .. \hat Z_5
         \rangle^4}{\langle  \hat Y \hat Z_1\hat Z_2\hat Z_3\hat Z_4 \rangle \dots \langle
         \hat Y\hat Z_5\hat Z_1\hat Z_2\hat Z_3\rangle}\notag\\
    & \qquad \qquad \downarrow\notag\\& \frac{  \langle \hat Y d^4\hat
                                        Y\rangle \langle  \hat Z_1 .. \hat Z_5
                                        \rangle^4}{\langle  \hat Y
                                        \hat Z_1\hat Z_2\hat Z_3\hat Z_4 \rangle \dots
                                        \langle \hat Y\hat Z_5\hat Z_1\hat Z_2\hat Z_3\rangle}
    \end{align}

    Here in the first line we used that under  the
    replacement~\eqref{eq:29} 
    \begin{align}
      \langle X_1 X_2 X_3X_4X_5 \rangle = \langle Y_1 ..Y_5 \hat
      Z_1..\hat Z_5  \rangle
      \prod_{i=1}^5(\tau_i.\sigma_{i+1})^{-1}\ .
    \end{align}
 as well as~\eqref{eq:30}
    \begin{align}
      \langle Y X_i X_j \rangle = \langle Y \hat Z_{i{-}1} \hat Z_i
      \hat Z_{j{-}1} \hat Z_j\rangle \times (\tau_{i{-}1}.\sigma_i
      \,\tau_{j{-}1}.\sigma_j)^{-1}\ .
    \end{align}
Finally in the last step we performed the reduction by simply removing
the total  derivatives involving the frozen variables
$d\sigma_i,d\tau_i$ which appear in an invariant measure.

The final result is precisely the five-point NMHV amplituhedron form.

We note that it is easier to consider
the functions without the measures (which are also much closer to the
actual correlator/amplitude expressions). Also it is then easier to
make particular choices for the $\sigma_i, \tau_i$ for example
$\sigma_i^\alpha=(0,1)$, $\tau_i^\alpha=(1,0)$. Then the lightlike limit takes the
correlahedron expression  $g_{n;k}(Y,X_i)$ to the amplitude expression
$a_{n;k}(\hat Y,\hat Z_i)$ via the simple replacements, implementing
the action of freezing and projecting~(\ref{eq:42})
\begin{align}
  g_{n;k}(Y,X_i)\quad   \xrightarrow[ X_{i2}\rightarrow -\hat
  Z_{i},\,X_{i1}\rightarrow{-}\hat Z_{i-1}{-}Y_{i-1}]{
\quad Y_i = X_{i2} - X_{i{+}1\,1} (i=1\dots n),\  Y_{n{+}i}=\hat Y_i
  (i=1..k)\quad } \quad   a_{n;k}(\hat Y,\hat Z_i)\ .\label{eq:33}
\end{align}

We give a highly non-trivial example of this reduction procedure in
appendix~\ref{sec:lightlike-limit-nmhv}. There we reduce the
correlator $G_{6;1}$ given by the lengthy expression in~(\ref{AB})
to the corresponding NMHV 6 point amplitude.

\subsection{The non-maximal limit geometrically}

\label{sec:non-maximal-limit}

The maximal, $n$-point, lightlike limit described above reduces the
correlahedron, which lives in $\Gr(n{+}k,4{+}n{+}k)$ to $\Gr(k,4{+}k)$ by partial
freezing and 
projecting from $Y$. Physically it
reduces the $n$-point, Grassmann degree $k$ correlator $G_{n;k}$ to
the (square of the) tree level $n$-point   N${}^k$MHV
amplitude. However it is also possible to consider lightlike limits of
fewer points, $n'<n$. In this limit the correlator reduces to higher loop amplitudes, specifically the
$(n{-}n')$-loop, N${}^{k'}$MHV amplitude, $A_{n';k'}^{(n-n')}$ where 
\begin{align}
k'= k-n+n'\ .\label{eq:4}
\end{align}

As in section~\ref{sec:maxim-lightl-limit}, the light like limit is
taken by setting $\langle Y X_{i'} X_{i'+1}\rangle=0$ so that we are
freezing  $Y$ to intersect the $n'$ 4-planes,  $X_1\wedge X_2$, $X_2
\wedge X_3, \dots, X_{n'}\wedge X_1$.  We then project through these
$n'$ intersection points, but here we {\em also project through the
  $n-n'$ additional 2-planes $X_{n'{+}1}, \dots X_{n}$}.  This extra
step corresponds to integrating out the supersymmetric parts of $X_i$
for $i>n'$ leaving a space-time integrand.

The concrete description of this procedure starts as in
the maximal case: the imposition of $\langle Y X_i X_{i+1}\rangle=0$ allows us to  gauge fix (freeze) the first $n'$
components of $Y$ to take the form
\begin{align}
   Y_i = \sigma_i^\alpha X_{i\alpha} - \tau_i^\alpha X_{i{+}1\,\alpha}  \qquad i=1\dots n'\ \text{(cyclically)}.\label{eq:6}
\end{align}
However, in the non-maximal case we further
 gauge fix the next $2n-2n'$ components of the $Y$ matrix as follows:
 \begin{align}
   Y_{n'{+}1}&= \cL_{1\,1}+\sigma_{n'{+}1}^\alpha X_{n'{+}1\,\alpha}\, \qquad&   Y_{n'{+}2}&=
   \cL_{1\,2}+\sigma_{n'{+}2}^\alpha X_{n'{+}1\,\alpha}\,\notag\\
   Y_{n'{+}3}&= \cL_{2\,1}+\sigma_{n'{+}3}^\alpha X_{n'{+}2\,\alpha}\, \qquad&   Y_{n'{+}4}&=
   \cL_{2\,2}+\sigma_{n'{+}4}^\alpha X_{n'{+}2\,\alpha}\,\notag\\
&\dots&&\dots\notag\\
 Y_{2n{-}n'{-}1}&= \cL_{n{-}n'\,1}+\sigma_{2n{-}n'{-}1}^\alpha X_{n\,\alpha}\, \qquad&   Y_{2n{-}n'}&=
   \cL_{n{-}n'\,2}+\sigma_{2n{-}n'}^\alpha X_{n\,\alpha}\,,\label{eq:5}
 \end{align}
where the $\cL_{i\alpha}$ are transverse to all the $X_{n'{+}i\,\alpha}$
and  $Y_{i'}, i=1..n'$. 
Note that
\eqref{eq:5} is not a restriction on the hyperplane $Y$ but merely on
a choice of basis for $Y$; we can always choose a $GL(n+k)$
transformation to obtain~\eqref{eq:5},  unlike ~\eqref{eq:6} which
follows from the freezing $Y$ to the boundary of the space. 
 
 There will be $n{+}k - ( 2n{-}n') = k'$ components of $Y$ remaining and  we denote these by
 \begin{align}
   Y_{2n-n'+p'} = \hat Y_{p'}\qquad p' =1..k'\ \label{eq:8}
 \end{align}
 and we also insist that they are transverse  to both $X_{n'{+}i\,\alpha}$ and $Y_i, i=1..n'$ using $GL(n+k)$.

 To make the above statements precise we can choose a basis (but the final answer will be basis independent)
for $\C^{k{+}n{+}4}$ given by 
\begin{align}
\text{basis} = \Big\{Y_1, \dots, Y_{n'}, X_{n'{+}1\,1},X_{n'{+}1\,2},\dots ,X_{n\,1},X_{n\,2},   e_1,\dots, e_{k'+4}\Big\}\ ,
\end{align}
 where $e_1,\dots e_{k'+4}$
are any $k'+4$ vectors such that this yields an independent
basis.\footnote{As in the maximal case, geometrically the span of the $e_i$ gives a hyperplane onto
which we are projecting the quotient.}
The projection then corresponds simply to setting to zero the first $2n-n'$ components of any vector in this basis
\begin{align}
X_{i\alpha} \rightarrow \hat X_{i\alpha} \quad \text{where} \quad
 \hat X_{i\alpha}^\cA = \left\{
  \begin{array}{ll}
0 \quad &\cA =1,\dots,2n-n'\\
     X_{i\alpha}^\cA \quad &\cA=2n{-}n'{+}1,\dots, n{+}k{+}4
  \end{array}
\right.\ .
\end{align}
We will have reduced brackets on the projected $k'+4$ dimensional space spanned by
$\{e_1,\dots,e_{k'+4}\}$
\begin{align}
  \langle \hat \cX  \rangle :=   \langle Y_1 \dots Y_n X_{n'+1}..X_{n}  \cX  \rangle\ .
\end{align}
Here $\cX$ represents any collection of $k'+4$ independent vectors, and
$\hat \cX$ the same vectors projected onto the hyperplane. 

As in the maximal case we define
\begin{align}
  Z_i:=\sigma_i.X_i &= \tau_i.X_{i+1} + Y_i  \quad i=1..n'\ \label{eq:28b}
\end{align}
and after projection this implies
\begin{align}
  \hat Z_i:=\sigma_i.\hat X_i = \tau_i.\hat X_{i+1}\quad i=1..n'\  ,
\end{align}
the projected planes $\hat X_i$  intersect each other
consecutively at $\hat Z_i$ in
the projected space.

If we choose
coordinates such that $\tau_{i-1}.\sigma_i>0$ for all $i=1..n'$ and
$\sigma_{n'+2a-1}.\sigma_{n'+2a}>0$ for all $a=1..n$ (ie  make a  choice of orientation for the
projection planes) then the correlahedron region  becomes
\begin{align}
  \label{eq:9}
  \langle Y X_i X_j\rangle>0 \rightarrow \left\{
  \begin{array}{l}
\langle   \hat Y \hat Z_{i-1} \hat Z_i \hat Z_{j-1}\hat Z_j  \rangle>0 \qquad i,j \in
    \{1,..,n'\}\\
\langle  \hat Y\cL_{i-n'} \hat Z_{j-1}\hat Z_j  \rangle>0 \qquad j \in
    \{1,..,n'\}\quad i \in
    \{n'{+}1,..,n\}\\
\langle   \hat Y \cL_{j-n'} \hat Z_{i-1} \hat Z_i   \rangle>0 \qquad i \in
    \{1,..,n'\}\quad j \in
    \{n'{+}1,..,n\}\\
\langle  \hat Y \cL_{i{-}n'}\cL_{j{-}n'} \rangle>0
  \qquad i,j \in
    \{n'{+}1,..,n\}\\
  \end{array}
\right.
\end{align}
This region is precisely the loop level squared amplituhedron
region~(\ref{eq:38}).

\subsection{The non-maximal limit on the hedron expressions}

\label{sec:non-maximal-limit-1}

As in  the maximal case, the ``freeze and project'' procedure can be applied
directly on the correlahedron form also for the non-maximal
limit. The procedure in the maximal case was given in
section~\ref{sec:lightl-limit-corr} and the non-maximal case is very similar.
When we freeze $Y$ as in~\eqref{eq:5} we get
\begin{align}
      \langle Y d^4 Y_i\rangle = \langle Y X_{i} X_{i+1} \rangle
      d^2\sigma_i d^2\tau_i \qquad i=1..n'\ .
    \end{align}
We perform the projection on the differential form by the map
\begin{align}
X_{i\alpha} \rightarrow \frac{-\tau_{i{-}1\, \alpha}\hat Z_i +\sigma_{i\,
  \alpha}(\hat Z_{i-1}+Y_{i-1})}{\tau_{i-1}.\sigma_i}\quad i=1..n'\ .\label{eq:29b}
\end{align}
 
The correlahedron form is then reduced to the amplituhedron form
by setting $Y$  as in~\eqref{eq:5} and 
$X_{i\alpha}$ to~\eqref{eq:29b}  and finally leaving out the
$\sigma,\tau$ dependent pieces:
\begin{align}
&  \prod_{i=1}^{n+k}  \langle Y d^4Y_i \rangle \times \frac{g_{n;k}(
  Y,  X_i) }{  \prod_{i=1}^{n'} \langle Y X_i
  X_{i+1}\rangle   } \notag\\
&\rightarrow \left(\prod_{i=1}^{n'} {d^2 \sigma_i
  d^2\tau_i}\right) \left(\prod_{i=1}^{n-n'} \prod_{\alpha=1}^2d^2 \sigma_{n'+2i-2+\alpha}
  \langle Y  X_{n'+i} d^2\cL_{i\alpha}\rangle  \right)\left(\prod_{i=1}^{k'}\langle \hat Y d^4\hat Y_i
  \rangle\right) g_{n;k}\Big(Y, X_{i\alpha}\Big)\notag\\
&\rightarrow\left(\!\prod_{i=1}^{n} \frac{d^2 \sigma_i
  d^2\tau_i}{(\tau_{i-1}.\sigma_i)^2} \! \right) \!\!\left(\prod_{i=1}^{n-n'} \frac{d^2 \sigma_{n'+2i-1}
d^2 \sigma_{n'+2i}  }{(\sigma_{n'+2i-1}.\sigma_{n'+2i})^2}
  \prod_{\alpha=1}^2 \langle Y_1..Y_{n'}X_{n'+1}..X_{n} \hat Y \cL_i d^2\cL_{i\alpha}\rangle\!\right)\!\! \left(\!\prod_{i=1}^{k'}\langle \hat Y d^4\hat Y_i
  \rangle\!\right) a_{n;k}^{(n-n')}
\notag\\
&\rightarrow \left(\prod_{i=1}^{n-n'} \prod_{\alpha=1}^2 \langle \hat
  Y \cL_i d^2\cL_{i\alpha}\rangle\right) \left(\prod_{i=1}^{k'}\langle \hat Y d^4\hat Y_i
  \rangle\right) a_{n;k}^{(n-n')}(\hat Y,\hat Z_i,\cL_i)
\end{align}
We proceeded in three stages. To get the second line we replaced $Y$
with~\eqref{eq:5} and~\eqref{eq:8} to get the third line we replaced
$X_{i\alpha}$ with~\eqref{eq:29b} and defined 
$a_{n;k}^{(n-n')}(\hat Y,\hat Z_i,\cL_i)$ which should correspond to  the square of
  the amplitude.  
We also used that
\begin{align}
  \langle Y  X_{n'+i} d^2\cL_{i\alpha}\rangle =  \langle
  Y_1..Y_{n'}X_{n'+1}..X_{n} \hat Y \cL_i d^2\cL_{i\alpha}\rangle\ .
\end{align}
We claim that the precise
dependence on $\sigma,\tau$ always has the factorised
form of the third line ie
\begin{align}
&   g_{n;k}\Big(Y, X_{i\alpha} \Big) \notag\\
&\rightarrow \left(\prod_{i=1}^{n'} \frac{1}{(\tau_{i-1}.\sigma_i)^2} \right)
 \left(\prod_{i=1}^{n-n'} \frac{1}{(\sigma_{n'+2i-1}.\sigma_{n'+2i})^2} \right) a_{n;k}^{(n-n')}(\hat Y,\hat Z_i,\cL_i)\ ,\label{eq:31b}
\end{align}
which can be seen as a consequence of the duality.

Consider for example  the {\em four}-point light-like limit of the five-point correlahedron $G_{5;1}$. 
We have  $Y=Y_1\wedge \dots \wedge Y_6 \ \in\ \Gr(6,10)$ and we freeze
$Y_1,\dots,Y_4$ to  $Y_i = \sigma^\alpha X_{i\alpha} - \tau^\alpha
X_{i{+}1\,\alpha}$, as in~\eqref{eq:6}, leaving $Y_5,Y_6$ free, which
we gauge fix as $Y_5=
\cL_{1\,1}+\sigma_{5}^\alpha X_{5\,\alpha},\ Y_6=
\cL_{1\,2}+\sigma_6^\alpha X_{5\,\alpha}$~\eqref{eq:5}. The projection
means we replace~\eqref{eq:29b} $X_{i\alpha} \rightarrow \frac{-\tau_{i{-}1\, \alpha}\hat Z_i +\sigma_{i\,
  \alpha}(\hat Z_{i-1}+Y_{i-1})}{\tau_{i-1}.\sigma_i}\quad i=1..4 $. Then
  \begin{align}
 &\prod_{i=1}^{6} \langle Yd^4Y_i \rangle    \frac{   \langle X_1 X_2
   X_3 X_4 X_5 \rangle^4}{\langle YX_1X_2\rangle \dots \langle
   YX_4X_5\rangle}\notag\\ 
    \xrightarrow[\text{project} X]{\text{freeze $Y$}}
    &\left(\prod_{i=1}^4
                                      {\frac{d^2\sigma_id^2\tau_i}{(\tau_{i-1}.\sigma_i)^2}}\right)\frac{d^2\sigma_5d^2\sigma_6}{(\sigma_{5}.\sigma_6)^2}
                                      \frac{   \langle Y_1..Y_4 
                                                     X_5 \cL_1 d^2\cL_{1\,1}\rangle \langle Y_1..Y_4 
                                                     X_5 \cL_1 d^2\cL_{1\,2}\rangle
                                      \langle Y_1..Y_4 X_5 \hat Z_1 .. \hat Z_4
                                      \rangle^4}{\langle Y_1..Y_4 X_5
                                      \hat Z_1\hat Z_2\hat Z_3\hat Z_4\rangle^2 \prod_{i=1}^4\langle
                                      Y_1..Y_4 X_5 \cL_1 \hat Z_{i-1}\hat Z_i
                                      \rangle }\notag
    \\=&\left(\prod_{i=1}^4 {\frac{d^2\sigma_id^2\tau_i}{(\tau_{i-1}.\sigma_i)^2}}\right)\frac{d^2\sigma_5d^2\sigma_6}{(\sigma_{5}.\sigma_6)^2}  \frac{\langle \cL_1 d^2\cL_{1\,1}\rangle \langle  \cL_1 d^2\cL_{1\,2}\rangle
         \langle  \hat Z_1 .. \hat Z_4 \rangle^2}{\prod_{i=1}^4\langle  \cL_1 \hat Z_{i-1}\hat Z_i \rangle}\notag\\
    & \qquad \qquad \downarrow\notag\\& \frac{ \langle \cL_1
                                        d^2\cL_{1\,1}\rangle
                                        \langle  \cL_1
                                        d^2\cL_{1\,2}\rangle
                                        \langle \hat Z_1 .. \hat Z_4
                                        \rangle^2}{\langle  \cL_1\hat Z_1\hat Z_2 \rangle \langle  \cL_1\hat Z_2\hat Z_3 \rangle\langle  \cL_1\hat Z_3\hat Z_4 \rangle\langle  \cL_1\hat Z_4\hat Z_1 \rangle}\label{eq:32}
    \end{align}
Here we used 
   \begin{align}
      \langle Y d^4 Y_i\rangle &= \langle Y X_{i} X_{i+1} \rangle
     d^2\sigma_i d^2\tau_i \qquad  i=1..4 \ \text{(cyclically)}\notag\\
 \langle Y d^4 Y_5\rangle  &= d^2\sigma_5 \langle Y_1..Y_4 
                                                     X_5 \cL_1 d^2\cL_{1\,1}
                                                     \rangle\notag\\
 \langle Y d^4 Y_6\rangle  &= d^2\sigma_6 \langle Y_1..Y_4 
                                                     X_5 \cL_1 d^2\cL_{1\,2}
                                                     \rangle
    \end{align}
and notice that $\langle Y X_{i} X_{i+1} \rangle$
cancels four terms of the denominator.
    Also in the first line we used
    \begin{align}
      \langle X_1 X_2 X_3X_4X_5 \rangle \rightarrow \langle Y_1 ..Y_4 X_5
      \hat Z_1..\hat Z_4  \rangle   \prod_{i=1}^4(\tau_i.\sigma_{i+1})^{-1}
    \end{align}
    after the projection~\eqref{eq:29b}. 

The result~\eqref{eq:32} is precisely the one-loop four-point
amplituhedron 
form.

Just as in the maximal case we again note that it is easier to consider
the functions without the measures (which are also much closer to the
actual correlator/amplitude expressions). Also we can then 
make particular choices for the $\sigma_i, \tau_i$ for example
$\sigma_i=(1,0)$, $\tau_i=(0,1)$. Then the lightlike limit takes the
correlahedron expression  $g_{n;k}(Y,X_i)$ to the amplitude expression
$g(\hat Y,\hat Z_i)$ via~\eqref{eq:31b}
\begin{align}
\label{eq:43}
 g_{n;k}(Y,X_i)  \xrightarrow[ X_{i1}\rightarrow \hat
  Z_{i-1}{+}Y_{i-1},\,X_{i2}\rightarrow{-}\hat Z_i,(i=1..n')\ ,\ Y_{2n-n'{+}i}=\hat Y_i
  (i=1..k')]{
\quad Y_i = X_{i1} - X_{i{+}1\,2} (i=1..n'),\
  Y_{n'{+}2i-2+\alpha}=
                        \cL_{1\,\alpha}+\sigma_{n'{+}2i-2+\alpha}^\alpha
                        X_{n'{+}i\,\alpha},\ (i=1..n{-}n')\quad } \quad   a_{n;k}^{(n-n')}(\hat Y,\hat Z_i,\cL_i)\ .
\end{align}

We give a highly non-trivial example in the non-maximal lightlike
limit case in the appendix where we consider the five-point lightlike
limit of the six point correlator $G_{6;1}$ and show that it correctly 
reproduces the five-point one-loop amplitude.

\section{Hedron expressions from hedron geometry}

\label{sec:computations}

We have introduced the correlahedron as a geometric object in
$\Gr(k{+}n,k{+}n{+}4)$.  We have also shown how to translate explicit
expressions for the correlator in analytic superspace to invariant
differential forms on $\Gr(k{+}n,k{+}n{+}4)$. The question we wish to
address in this 
section is the direct relation between the correlahedron geometry and
the corresponding  differential form. We will only give a tentative
answer 
to this question here, leaving further developments to future work. 
Working towards this however we first concentrate on the analogous issue for
the squared amplituhedron.  For the amplituhedron itself a
prescription for 
obtaining the amplitude from the geometry was defined
in~\cite{Arkani-Hamed:2013jha}. To obtain the amplitude from the
amplituhedron it was conjectured that one takes the volume form with no 
divergences inside the amplituhedron and logarithmic
divergences on the boundary. This defines a volume form on
amplituhedron space which is equivalent to the bosonised amplitude.
We take exactly the same prescription here for the squared
amplituhedron. Furthermore, due to the more explicit description of
the squared amplituhedron, we are able to give a simple computerisable
algorithm (via cylindrical decomposition)  for obtaining this volume
form.

\subsection{Practical algorithm for obtaining the hedron  form  from
  the 
hedron region}

The amplitu-/correla-hedron is described geometrically as a subspace of a
Grassmannian space. In order to relate this to an amplitude or
correlator one has to obtain        a 
differential  form from this geometry. 
For the squared amplituhedron this is the unique form which has logarithmic
divergences on the boundary of the amplituhedron space and no 
divergences inside the space. Here we describe a simple algorithm for obtaining the form from the region.

The first step is to obtain  a
cylindrical decomposition of the region. A cylindrical decomposition of any subset of $R^n$  describes it as  a union
of regions with the form
\begin{align}
  \left\{
  \begin{array}{lrl}
  &   a<&x_1<b,\\
   &  a(x_1)<&x_2<b(x_1),\\
    (x_1,\dots ,x_n):\qquad  & a(x_1,x_2)<&x_3<b(x_1,x_2),\\ &\dots,\\
  &    a(x_1,..,x_{n-1})<&x_n<b(x_1,
..,x_{n-1})
  \end{array}
\right\}\ ,
\end{align}
ie each variable is restricted to an interval which depends on the
previous variables. 

This is exactly the description of a region one
needs 
to perform an integration over the region as a multiple integral. 
Here however instead of integrating over this region one assigns a differential form
to it by assigning to each inequality a dlog:
\begin{align}
a(x_1,..,x_{i{-}1})<x_i<b(x_1,..,x_{i{-}1})  &\quad \rightarrow \quad d
                                           \log \left(
                                               \frac{x_i-b(x_1,..,x_{i{-}1})}{x_i-a(x_1,..,x_{i{-}1})}\right)\
                                               \label{eq:15}
\end{align}
thus yielding the $n$-form
\begin{align}
\prod_{i=1}^n  \frac{ dx_i \Big(b(x_1,..x_{i{-}1})-a(x_1,..x_{i{-}1})\Big)}{\Big(x_i-b(x_1,..x_{i{-}1})\Big)\Big(x_i-a(x_1,..x_{i{-}1})\Big)}\ .
\end{align}
One then simply adds together the contributions from each region.
This gives a form with log divergences on each boundary and
no divergences inside (as long as the original region is convex).

We here describe this process through the  simplest example. We consider the case of a triangle in $P^2$ with vertices $Z_1,Z_2,Z_3$. We give them inhomogeneous coordinates $Z_i=(x_i,y_i,1)$. The region (inside of the triangle) is the space of $Y \in P^2 $ such that
\begin{align}
  \langle Y Z_1 Z_2 \rangle > 0, \quad    \langle Y Z_2 Z_3 \rangle > 0, \quad    \langle Y Z_3 Z_1 \rangle > 0 \ .
\end{align}
Let us also  give $Y$ inhomogeneous coordinates $Y=(x,y,1)$ the region
becomes
\begin{align}
\begin{tikzpicture}
  \coordinate[label={$(x_3,y_3)$}] (three) at (1.3,1.9);
  \coordinate[label=left:{$(x_1,y_1)$}] (one) at (-1.5,-2.3) {};
  \coordinate[label=right:{$(x_2,y_2)$}] (two) at (2.7,-2.9) {};
  \coordinate (four) at (1.3,0);
\draw (one) -- (two) -- (three) -- (one);
\draw[dashed] (three) -- (intersection of three--four and one--two);
\end{tikzpicture}
\end{align}
and can be written as the sum of two regions
\begin{align}
  \frac{ x y_1 - x_2 y_1 - x y_2 + x_1 y_2}{x_1 - x_2} < y <  \frac{x y_1 - x_3 y_1 - x y_3 + x_1 y_3}{x_1 - x_3} &\text{ and } x_1<x<x_3\notag\\
    \frac{ x y_1 - x_2 y_1 - x y_2 + x_1 y_2}{x_1 - x_2} < y <  \frac{x y_2 - x_3 y_2 - x y_3 + x_2 y_3}{x_2 - x_3}&\text{ and } x_3<x<x_2\ .
\end{align}

So the differential form corresponding to the above region becomes

\small
\begin{align}&
  d \log\left( \frac{y - \frac{x y_1 - x_3 y_1 - x y_3 + x_1 y_3}{x_1 - x_3} } {y- \frac{x y_1 - x_2 y_1 - x y_2 + x_1 y_2}{x_1 - x_2}}  \right) \wedge d\log\left( \frac{x-x_3}{x-x_1}  \right) + d \log\left( \frac{y - \frac{x y_2 - x_3 y_2 - x y_3 + x_2 y_3}{x_2 - x_3} } {y- \frac{x y_1 - x_2 y_1 - x y_2 + x_1 y_2}{x_1 - x_2}}  \right) \wedge d\log\left( \frac{x-x_2}{x-x_3}  \right)\notag\\
  &=\frac{dx dy\left(x_2 y_1{-}x_3 y_1{-}x_1 y_2{+}x_3 y_2{+}x_1 y_3{-}x_2
   y_3\right){}^2}{\left(x_1 y{-}x_1 y_2{-}x_2 y{-}x y_1{+}x_2 y_1{+}x
   y_2\right) \left(x_1y{-}x_1 y_3{-}x_3 y{-}x y_1{+}x_3 y_1{+}x
   y_3\right) \left(x_2y{-}x_2 y_3{-}x_3 y{-}x y_2{+}x_3 y_2{+}x
    y_3\right)}
    \notag\\
 & = \frac{\langle Y d^2Y \rangle \langle Z_1Z_2Z_3 \rangle^2}{\langle YZ_1Z_2\rangle\langle YZ_2Z_3\rangle\langle YZ_3Z_1\rangle }\ .\label{eq:31}
\end{align}
\normalsize

To get the second line we simply applied the differential and
factorised the result and to obtain the third line we simply rewrote
back in homogeneous coordinates. The final result is the  2-form associated with the triangle (see eg~\cite{Arkani-Hamed:2013jha}.)

The above method can be applied more generally and importantly can be 
simply implemented using a computer algebra programme (for numeric external vertices at least).
For example in {\tt mathematica} one can apply the command {\tt{CylindricalDecomposition[]}} to convert any set of inequalities into the form of a sum of regions upon which we can implement the simple rule~\eqref{eq:15}.

In the next two subsections we illustrate this procedure  in a number
of tree and loop examples.

\subsection{Tree level squared amplituhedron  examples}

\subsubsection{Five-point NMHV amplitude}

We begin with the simplest physical example, 5 point tree-level. The external
data is given by five points, $Z_1, \dots Z_5$, in $P^4$ and we obtain the
geometrical amplituhedron squared region as  $Y \in P^4$ subject to 
\begin{align}
  \langle YZ_i Z_{i+1} Z_{i+2} Z_{i+3} \rangle >0 \ .\label{eq:12}
\end{align}
This region arises directly from~\eqref{eq:9}.

To make this concrete introduce coordinates  by $Y=y_1Z_1+y_2Z_2+y_3Z_3+y_4Z_4+Z_5$ so the region becomes simply
\begin{align}
  y_1,y_2,y_3,y_4 >0
\end{align}
and the corresponding differential form is then trivially
\begin{align}
\frac{  dy_1dy_2dy_3dy_4}{y_1y_2y_3y_4}\ .\label{eq:13}
\end{align}
Finally we can covariantise this differential form to the coordinate independent  form
\begin{align}
\frac{  \langle Yd^4 Y \rangle  \langle12345\rangle^4}{\langle Y1234\rangle \langle Y2345\rangle\langle Y3451\rangle\langle Y4512\rangle\langle Y5123\rangle}\ .
\end{align}

This correctly reproduces the known amplitude (as a form in
amplituhedron space).
Note that in this case the description is entirely equivalent to the
amplituhedron itself (as compared to the squared amplituhedron). We note here that if one instead had a different orientation for one of the $X_i=Z_{i-1}\wedge Z_i$ then although the region~\eqref{eq:12} would be different, the resulting differential form would be the same. For example imagine that instead of $Z_4\wedge Z_5$ we had the reverse order $Z_5\wedge Z_4$ with all other edges having the same orientation.  Then the corresponding region in $P^4$ would be defined by $\langle YZ_2Z_3 Z_5 Z_4\rangle>0$ and $\langle YZ_5Z_4Z_1Z_2\rangle>0$, but with all other inequalities the same. In coordinates we would have $x_2,x_4>0$ as before, but this time $x_1 x_3<0$. However the resulting form~\eqref{eq:13} is the same.

\subsubsection{Six-point NNMHV }

\label{sec:six-point-nnmhv}

The next case, six-point NNMHV is more interesting. We consider the external points $Z_1,..,Z_6 \in P^5$ and the subspace of the Grassmannian of 2-planes $Y=Y_1\wedge Y_2 \in \Gr(2,6)$ defined by the inequalities
\begin{align}
  \langle Y i\,i{+}1\,j\,j{+}1 \rangle >0 \ .\label{eq:14}
\end{align}
Note that this is a weaker requirement than that of the amplituhedron which requires all ordered minors in the matrix defining  $Y$ to be positive (ie it requires additional constraints such as $\langle Y1235 \rangle>0$). 

Again then to obtain the differential form from this, we first coordinatise the $Y$s, letting
\begin{align}
 \left( \begin{array}{c}
  Y_1\\Y_2
  \end{array}
\right) = \left(
\begin{array}{cccccc}
 1 & a & b & 0 & c & d \\
 0 & e & f & 1 & g & h \\
\end{array}
\right)\left( \begin{array}{c}
  Z_1\\\vdots\\Z_6
  \end{array}
\right)
\end{align}
In these coordinates the inequalities~\eqref{eq:14} become
\begin{align}
  e > 0,  h > 0, b e - a f > 0, b > 0, -d f + b h > 0, 
 c e - a g > 0, c > 0, -d g + c h > 0\ .
\end{align}
Performing a cylindrical decomposition of these inequalities in the order $e, h, c, b, g, f, a, d$ (which seems to give the simplest result - the final answer for the differential form does not depend on this order) gives a description of the region as
{\scriptsize \begin{align}
  &e>0\land h>0\land c>0\land b>0\land \notag\\
  &\left(\left(g<0\land \left(\left(f<\frac{b g}{c}\land a>\frac{b e}{f}\land d>\frac{b
   h}{f}\right)\lor \left(\frac{b g}{c}<f<0\land a>\frac{c e}{g}\land d>\frac{c h}{g}\right)\lor \left(f>0\land \frac{c
   e}{g}<a<\frac{b e}{f}\land \frac{c h}{g}<d<\frac{b h}{f}\right)\right)\right)\right.\notag\\&\left.\lor \left(g>0\land \left(\left(f<0\land
   \frac{b e}{f}<a<\frac{c e}{g}\land \frac{b h}{f}<d<\frac{c h}{g}\right)\lor \left(0<f<\frac{b g}{c}\land a<\frac{c
   e}{g}\land d<\frac{c h}{g}\right)\lor \left(f>\frac{b g}{c}\land a<\frac{b e}{f}\land d<\frac{b
   h}{f}\right)\right)\right)\right)
\end{align}
}
which on performing the replacement~\eqref{eq:15} gives the remarkably simple differential form
\begin{align}
 \frac{2 (-a b g h-a c f h+a d f g+3 b c e h-b d e g-c d e f)}{b c e h (b e-a f) (c e-a g) (b h-d f) (c h-d g)}da \wedge\dots \wedge dh \ .
\end{align}
This lifts into the covariant form
\begin{align}
\frac{ 2 \langle Y d^4 Y_1\rangle  \langle Y d^4 Y_2\rangle \Big( \langle Y 3456 \rangle\langle Y 2361 \rangle\langle Y 1245 \rangle + \text{ cyclic} \Big)\langle123456\rangle^4}{\langle Y1245\rangle\langle Y 2356\rangle\langle Y3461\rangle\prod_{i=1}^6 \langle Y i\,i{+}1\,i{+}2\,i{+}3\rangle}\ .\label{eq:16}
\end{align}

We will discuss the interpretation of this in a moment but first let
us check what happens if we switch the orientation of one of the
edges. Specifically, we replace the edge $Z_6 \wedge Z_1$ with $Z_1
\wedge Z_6$. This swaps the inequality of three of the brackets
in~\eqref{eq:14}: $\langle Y 2316 \rangle>0,\langle Y 3416
\rangle>0,\langle Y 4516 \rangle>0$. But unlike the NMHV case (where this made no difference to the final differential form) here these swaps of signs make an enormous difference.  

Proceeding as in the previous case, with the same coordinates, the inequalities become
\begin{align}
  {e > 0,  h > 0, -b e + a f > 0, -c e + a g > 0, 
 b > 0, -d f + b h > 0, -c > 0, -d g + c h > 0}
\end{align}
and a cylindrical decomposition becomes even simpler:
\begin{align}
  &e>0\land h>0\land c<0\land b>0\land \notag\\
  &\left(\left(g<0\land f<0\land a<\frac{b e}{f}\land d>\frac{c h}{g}\right)\lor
   \left(g>0\land f>0\land a>\frac{b e}{f}\land d<\frac{c h}{g}\right)\right)
\end{align}
yielding the differential form
\begin{align}
  \frac{2 da\wedge \dots \wedge dh}{b c e h (b e-a f) (c h-d g)}\ ,
\end{align}
which in turn covariantises to
\begin{align}
\frac{2  \langle Y d^4 Y_1\rangle  \langle Y d^4 Y_2\rangle \langle123456\rangle^4}{\prod_{i=1}^6 \langle Y i\,i{+}1\,i{+}2\,i{+}3\rangle}\ .
\end{align}

So in this case we thus obtain two different answers depending on the
orientation of the edges. In fact remarkably both answers have a
physical meaning. The result arising from  the cyclic choice of
orientation~\eqref{eq:16}  corresponds to  the square of the NMHV
amplitude $(NMHV_6)^2$ whereas the  result from the non-cyclic
ordering yields (twice ) the NNMHV amplituhedron. The lightlike limit
of the correlahedron yields the sum of these two terms. Furthermore we
find that all other choices of orientations for the edges yield the same results: an odd number of edge flips yields the amplituhedron, an even number yields (NMHV)${}^2$.
Given this result it is natural to conjecture that in all cases 
the correlahedron is the average of all possible orientations of the edges.

\subsubsection{Seven-point N\texorpdfstring{${}^3$}{3}MHV}

As a final tree-level example we consider the
 seven-point N${}^3$MHV amplitude,  described as a subspace of $\Gr(3,7)$ with the external data $Z_i$ living in $P^6$. The subspace is defined as the set $Y=Y_1\wedge Y_2\wedge Y_3\in \Gr(3,7)$ such that
  \begin{align}
  \langle Y i\,i{+}1\,j\,j{+}1 \rangle >0\ .\label{eq:17}
\end{align}
Employing the same procedure as previously, we coordinatise $\Gr(3,7)$ as
\begin{align}
 \left( \begin{array}{c}
  Y_1\\Y_2\\Y_3
  \end{array}
\right) = \left(
\begin{array}{ccccccc}
 1 & a & b & 0 & c & d & 0 \\
 0 & e & f & 1 & g & h & 0 \\
 0 & i & j & 0 & k & l & 1 \\
\end{array}
\right)\left( \begin{array}{c}
  Z_1\\\vdots\\Z_7
  \end{array}
\right)
\end{align}
and then perform a cylindrical decomposition of the region~\eqref{eq:17} in these variables and then convert the result into a differential form according to~\eqref{eq:15}. Remarkably the result is precisely the lightlike limit of the 7 point correlator, or equivalently the square of the amplitude $2N^3MHV_7+2NMHV_7N^2MHV_7$. Explicitly it can be written in correlahedron space as
\begin{align}
  &{\langle Y d^4Y_1 \rangle  \langle Y d^4Y_2 \rangle  \langle Y d^4Y_3 \rangle\langle 1234567\rangle}^4\notag \\
  &\times \left(\frac{\langle Y7123\rangle }{\langle Y1234\rangle  \langle Y1267\rangle  \langle Y2345\rangle  \langle
   Y2356\rangle  \langle Y2367\rangle  \langle Y7134\rangle  \langle Y7145\rangle  \langle Y7156\rangle } + \dots \right)\ .\label{eq:18}
\end{align}
Here the first (displayed) term is the contribution of the N${}^3$MHV amplitude and the dots denote the contributions from the product amplitudes $NMHV_7N^2MHV_7$. The full expression is  most compactly written as the lightlike limit of the correlator,  which is the $S_7$ permutation of a single term. So the bit in brackets in~\eqref{eq:18} can be written
\begin{align}
\frac{  \lim_{x_{i\,i{+}1}^2\rightarrow 0} (x_{12}^4x_{34}^2x_{45}^2x_{56}^2x_{67}^2x_{37}^2 \ +\ \text{$S_7$ permutations})}{ \prod_{i=1}^7 x_{i\,i{+}2}^2x_{i\,i{+}3}^2  }  \qquad \text{with } \quad  x_{ij}^2 \rightarrow \langle Y\,i{-}1\,i\, j{-}1\,j \rangle\ .
\end{align}

It is remarkable that this expression arises very simply from the constraints~\eqref{eq:17}. Note that unlike the N${}^2$MHV case this single choice of edge orientation gives the full answer. Flipping the orientation of one or more of the edges yields exactly the same result in this case.

\subsection{Loop level squared amplituhedron examples}

As further illustration we now consider some loop level examples where
again the cylindrical decomposition procedure correctly reproduces the
squared amplitude.

\subsubsection{Four-point one-loop}

Here we have external twistors $Z_i \in P^3$ and the set of $\cL=\cL_1\wedge \cL_2 \in \Gr(2,4)$ subject to
\begin{align}
  \langle\cL 12 \rangle>0,\   \langle\cL 23 \rangle>0,\   \langle\cL 34 \rangle>0,\   \langle\cL 41 \rangle>0\ .
\end{align}
Putting coordinates for $\cL$ as
\begin{align}
 \left( \begin{array}{c}
  \cL_1\\\cL_2
  \end{array}
\right) = \left(
\begin{array}{cccccc}
 1 & 0 & a & b \\
 0 & 1 & c & d\\
\end{array}
\right)\left( \begin{array}{c}
  Z_1\\\vdots\\Z_4
  \end{array}
\right)
\end{align}
This yields the differential form
\begin{align}
  \frac{2 da \wedge db\wedge dc\wedge dd}{a d (a d-b c)}\ 
\end{align}
which lifts to
\begin{align}
\frac{ 2 \langle \cL d^2\cL_1 \rangle  \langle \cL d^2\cL_2 \rangle \langle1234\rangle^2}{\langle\cL12\rangle\langle\cL23\rangle\langle\cL34\rangle\langle\cL41\rangle}\ .
\end{align}

\subsubsection{Four-point two-loop}

Here we have external twistors $Z_i \in P^3$ and the set of $\cL=\cL_1\wedge \cL_2 \in \Gr(2,4)$ and $\cM=\cM_1\wedge \cM_2 \in \Gr(2,4)$  subject to
\begin{align}
  &\langle\cL 12 \rangle>0,\   \langle\cL 23 \rangle>0,\   \langle\cL 34 \rangle>0,\   \langle\cL 41 \rangle>0\notag\\
  &\langle\cM 12 \rangle>0,\   \langle\cM 23 \rangle>0,\   \langle\cM 34 \rangle>0,\   \langle\cM 41 \rangle>0,\ \langle\cL\cM\rangle>0\ .\label{eq:19}
\end{align}
Putting coordinates for $\cL$ and $\cM$ as
\begin{align}
 \left( \begin{array}{c}
  \cL_1\\\cL_2
  \end{array}
\right) = \left(
\begin{array}{cccccc}
 1 & 0 & a & b \\
 0 & 1 & c & d\\
\end{array}
\right)\left( \begin{array}{c}
  Z_1\\\vdots\\Z_4
  \end{array}
\right)\, , \qquad  \left( \begin{array}{c}
  \cM_1\\\cM_2
  \end{array}
\right) = \left(
\begin{array}{cccccc}
 1 & 0 & e & f \\
 0 & 1 & g & h\\
\end{array}
\right)\left( \begin{array}{c}
  Z_1\\\vdots\\Z_4
  \end{array}
\right)\, .\label{eq:20}
\end{align}
This yields the differential form (obtained as in the previous cases by writing the
inequalities~\eqref{eq:19} in terms of the coordinates~\eqref{eq:20},
obtaining a cylindrical decomposition of this region,  and then making
the replacement~\eqref{eq:15}) 
\begin{align}
  \frac{2 da \wedge \dots \wedge dh (2 a d-2 b c+b g+c f+2 e h-2 f
  g)}{a d e h (a d-b c) (e h-f g) (a d-a h-b c+b g+c f-d e+e h-f g)} \ . 
\end{align}
This lifts to
\begin{align}
&{ 2 \langle \cL d^2\cL_1 \rangle  \langle \cL d^2\cL_2 \rangle \langle
  \cM d^2\cM_1 \rangle  \langle \cM d^2\cM_2 \rangle
  \langle1234\rangle^3}\notag\\
&\times\left(
\frac{1}{\langle \mathcal{L}23\rangle  \langle \mathcal{L}34\rangle  \langle \mathcal{L}41\rangle  \langle \mathcal{M}12\rangle  \langle
   \mathcal{M}23\rangle  \langle \mathcal{M}41\rangle  \langle \mathcal{L}\mathcal{M}\rangle }+\frac{1}{\langle \mathcal{L}12\rangle  \langle
   \mathcal{L}34\rangle  \langle \mathcal{L}41\rangle  \langle \mathcal{M}12\rangle  \langle \mathcal{M}23\rangle  \langle \mathcal{M}34\rangle  \langle
   \mathcal{L}\mathcal{M}\rangle }\right.\notag\\&\quad \left.+\frac{1}{\langle \mathcal{L}12\rangle  \langle \mathcal{L}23\rangle  \langle \mathcal{L}34\rangle  \langle
   \mathcal{M}12\rangle  \langle \mathcal{M}41\rangle  \langle \mathcal{M}34\rangle  \langle \mathcal{L}\mathcal{M}\rangle }+\frac{1}{\langle
   \mathcal{L}12\rangle  \langle \mathcal{L}23\rangle  \langle \mathcal{L}41\rangle  \langle \mathcal{M}23\rangle  \langle \mathcal{M}41\rangle  \langle
   \mathcal{M}34\rangle  \langle \mathcal{L}\mathcal{M}\rangle }\right.\notag\\&\quad \left.+\frac{\langle1234\rangle}{\langle \mathcal{L}12\rangle  \langle \mathcal{L}23\rangle  \langle
   \mathcal{L}34\rangle  \langle \mathcal{L}41\rangle  \langle \mathcal{M}12\rangle  \langle \mathcal{M}23\rangle  \langle \mathcal{M}41\rangle  \langle
   \mathcal{M}34\rangle }\right)\ .
\end{align}
Here we recognise both the square of the one-loop amplitude (last
term) as well as the two loop amplitude (first four terms which are
all double boxes). The full expression is precisely the result of
taking the lightlike limit of the correlator, ie the square of the
four-point amplitude at second order in perturbation theory.

\subsubsection{Five-point one-loop}

Here we have external twistors $Z_i \in P^4$, the loop 2-plane
$\cL=\cL_1\wedge \cL_2 \in \Gr(2,5)$ as well as $Y \in P^4$. $Y$ and
$\cL$ satisfy the following inequalities
\begin{align}
  &\langle\cL Y 12 \rangle>0,\   \langle\cL Y 23 \rangle>0,\
    \langle\cL Y 34 \rangle>0,\   \langle\cL Y 45 \rangle>0,\   \langle\cL Y 51 \rangle>0\notag\\
  &\langle Y 1234 \rangle>0,\   \langle Y 2345 \rangle>0,\   \langle Y
    3451 \rangle>0,\   \langle Y 4512 \rangle>0,\ \langle Y5123\rangle>0\ .\label{eq:21}
\end{align}
Putting coordinates for $\cL$ and $Y$ as
\begin{align}
 \left( \begin{array}{c}
  \cL_1\\\cL_2
  \end{array}
\right) =\left(
\begin{array}{ccccc}
 1 & 0 & a & b & 0 \\
 0 & 1 & c & d & 0 \\
\end{array}
\right)\left( \begin{array}{c}
  Z_1\\\vdots\\Z_5
  \end{array}
\right)\, ,\qquad Y = \left(
\begin{array}{ccccc}
 e & f & 1 & g & h \\
\end{array}
\right)\left( \begin{array}{c}
  Z_1\\\vdots\\Z_5
  \end{array}
\right)\, ,
\end{align}
the inequalities~\eqref{eq:21} lead to the
differential form
\begin{align}
-\frac{2 a d e f-2 a e g-2 b c e f+b e-c f g+d f+2 g}{d e f g h (a d-b c) (a e+c f-1) (a d f-a g+b (-c) f+b)}da\wedge db\wedge\dots\wedge
  dh\ .
\end{align}
This lifts to the co-ordinate independent form
\begin{align}
& \frac{\langle \cL Y d^2\cL_1 \rangle  \langle \cL Y d^2\cL_2 \rangle
  \langle Yd^4Y\rangle  \langle12345\rangle^4}{\langle Y1234\rangle\langle Y2345\rangle\langle Y3451\rangle\langle Y4512\rangle\langle Y5123\rangle}\notag\\
&\times\left(
\frac{\langle 1234Y\rangle  \langle 2345Y\rangle }{\langle \mathcal{L}Y12\rangle  \langle \mathcal{L}Y23\rangle  \langle \mathcal{L}Y34\rangle 
   \langle \mathcal{L}Y45\rangle }+\frac{\langle 5134Y\rangle  \langle 2345Y\rangle }{\langle \mathcal{L}Y23\rangle  \langle \mathcal{L}Y34\rangle 
   \langle \mathcal{L}Y45\rangle  \langle \mathcal{L}Y51\rangle
  }\right.\notag\\&\quad \quad +\frac{\langle 1234Y\rangle  \langle 5123Y\rangle }{\langle \mathcal{L}Y12\rangle 
   \langle \mathcal{L}Y23\rangle  \langle \mathcal{L}Y34\rangle  \langle \mathcal{L}Y51\rangle }+\frac{\langle 1245Y\rangle  \langle 5123Y\rangle
   }{\langle \mathcal{L}Y12\rangle  \langle \mathcal{L}Y23\rangle
                                                                            \langle
                                                                            \mathcal{L}Y45\rangle
                                                                            \langle
                                                                           \mathcal{L}Y51\rangle                                                                            }\notag\\&\qquad                                                                           \qquad                                                                          \qquad \qquad                                                                 \left.+  \frac{\langle
   1245Y\rangle  \langle 5134Y\rangle }{\langle \mathcal{L}Y12\rangle  \langle \mathcal{L}Y34\rangle  \langle \mathcal{L}Y45\rangle  \langle
   \mathcal{L}Y51\rangle }\right)\ .
\end{align}
Here we recognise the sum of five box functions (which is the parity
even part of the one-loop amplitude) multiplied by the tree-level NMHV
amplitude. This is precisely what we expect: the square of the
superamplitude at first non-trivial order in both coupling and the
Grassmann odd 
variable expansion is
\begin{align}
  \label{eq:22}
  \left( \frac{A_\text{MHV}^{(0)}+A_\text{NMHV}^{(0)} +a
  A_\text{MHV}^{(1)}+a A_\text{NMHV}^{(1)} +
  \dots}{A_\text{MHV}^{(0)}}\right)^2|_{a^1,\eta^4}&=\frac{2 A_\text{MHV}^{(0)}
  A_\text{NMHV}^{(1)}  +A_\text{NMHV}^{(0)} A_\text{MHV}^{(1)}\notag}{\left(A_\text{MHV}^{(0)}\right)^2} \\&=2 
  \frac{A_\text{NMHV}^{(0)}}{A_\text{MHV}^{(0)}}\left(M_\text{NMHV}^{(1)}+M_\text{MHV}^{(1)}\right)\ ,\end{align}
where we define $M^{(\ell)}$ to be the loop level amplitude divided by tree-level
amplitude of the same helicity structure.

\subsection{Obtaining the correlator from the correlahedron}

We now arive at the question of how to obtain the correlator from the
correlahedron geometry. The obvious method is to attempt the same procedure successfully implemented above for the closely
related 
squared amplituhedron,
namely take the unique differential volume form on amplituhedron space
with log divergences on the boundary. There are two problems with
this. The first problem is purely practical in that the simplest example, the
five-point NMHV correlator 
$G_{5;1}$ is already far too high dimensional for the cylindrical
decomposition procedure to give a result (this procedure is doubly
exponential in the number of dimensions which is $4(k+n)= 24$ in this
case).
The second problem however is of a more serious nature  since it
suggests that such a naive implementation of the log divergence
criterion does not even apply straightforwardly in this case. The
problem is that the correlator apparently can have double poles on the
boundary, unlike the amplitude which always has single poles. We have
already seen examples of this feature, in for example
equation~\eqref{eq:42} where we see a denominator
$1/(\tau_{i-1}.\sigma_i)^2$. Such a double pole can not be obtained
naively from the cylindrical decomposition procedure.%
\footnote{ 
Note that log
divergence criterion for obtaining the differential form from the
geometry is valid also when we go to the boundary of the -hedron
space . In other words if we choose $Y$ to saturate  one or more of the
inequalities 
$\langle Y ..\rangle>0$ (so we pin ourselves to the
boundary of -hedron space ) then
implementing the cylindrical decomposition procedure on the remaining
inequalities/ variables yields the correct answer for the
residue of the expression in this limit.}

However since, as we saw in section~\ref{sec:lightlike-limit},  the
correlahedron geometry reduces to the amplituhedron by exactly the
same geometric procedure (freeze and project) 
as the corresponding differential form it would seem puzzling if the
procedure for obtaining the differential form from the geometry is
very different. The situation can be described by the following figure:

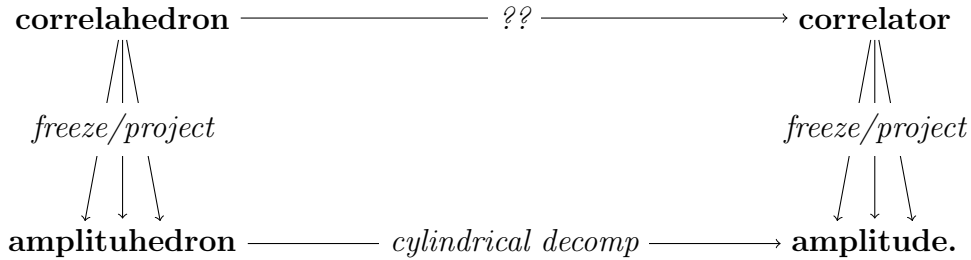
\begin{center}
  \begin{tikzpicture}
    \node (a) at (0,0) {\bf amplituhedron}; 
    \coordinate (a1) at (-.5,0.3); 
    \coordinate (a2) at (.5,0.3); 
    \coordinate (b1) at (9.5,0.3); 
    \coordinate (b2) at (10.5,0.3); 
    \node (b) at (10,0) {\bf      amplitude.}; 
    \node (c) at (0,3) {\bf correlahedron}; 
    \node (d) at
    (10,3) {\bf correlator}; 
    \draw[->] (a) -- (b) node [midway,
    fill=white] {{\it cylindrical decomp}}; 
   \draw[->] (c) --(a1);
  \draw[->] (c) --(a2);
   \draw[->] (c) --(a) node
    [midway, fill=white] {\it freeze/project}; 
 \draw[->] (d) --(b1);
 \draw[->] (d) --(b2);
    \draw[->] (d) --(b)
    node [midway, fill=white] {\it freeze/project}; 
    \draw[->]
    (c) -- (d) node [midway, fill=white] {\it ??};
  \end{tikzpicture}
\end{center}
(We have displayed multiple arrows from the correlahedron to the
amplituhedron to highlight the fact that one can take many different
limits to get many different amplitudes from the same correlator. )

A possible  resolution of this apparent puzzle arises
from 
a stronger  implementation of all the symmetries of the set up before
taking the cylindrical decomposition.

\subsubsection{Toy model reconsidered, implementing the full symmetry}

First consider again the simplest toy model case. There we consider
points 
$Y \in \Gr(1,3)$ inside the triangle formed by $Z_1,Z_2,Z_3 \in
\Gr(1,3)$. In the standard formulation we let $Y$ (with its two
degrees of 
freedom) vary fully inside the triangle. However in fact the global $GL(3)$
symmetry in this case allows one to completely fix $Y$, leaving no
degrees of freedom at all! To see this, first use $GL(3)$ to fix
$Z_1,Z_2,Z_3$ to the basis elements of $\R^3$ (using a projective
rescaling of each if
necessary) and set $Y=(y_1,y_2,1)$. Now consider the residual $GL(3)$
which leaves the external data $Z_i$ invariant. Since the $Z_i$ are
projective, the action of
the diagonal of $GL(3)$, diag$(a,b,c)$ can be removed by the
projective rescaling. 
On the other hand this residual $GL(3)$ acts as 
$Y\rightarrow (a/c y_1,b/c y_2,1)$. Thus by choosing $a,b,c$
appropriately 
we can use this to set for example $Y= (1,1,1)$.   We thus have no
degrees 
of freedom left at all if we implement the $GL(3)$ symmetry! 

In fact the triangle form~(\ref{eq:31}) can be completely
determined (up to an overall numerical constant) by these symmetries
alone. Indeed the  function of $Y,Z_1,Z_2,Z_3$ multiplying
$\langle{Yd^2Y}\rangle$ must be $GL(3)$ covariant, have weight zero in
the $Z_i$ and weight 3 in $Y$. The only possible function with these
properties is
\begin{align}
   \frac{\langle Y d^2Y \rangle \langle Z_1Z_2Z_3 \rangle^2}{\langle
  YZ_1Z_2\rangle\langle YZ_2Z_3\rangle\langle YZ_3Z_1\rangle }\ ,
\end{align}
the triangular form. So indeed this expression {\em can} be correctly
obtained from no 
degrees of freedom at all! 

\subsubsection{Amplituhedron squared reconsidered, implementing the
  full symmetry} 

The above example is a bit too trivial, so let us give another example.
Indeed one can reconsider the amplituhedron squared examples we looked
at in the previous subsections and implement the additional symmetry
in a similar way and show that the cylindrical 
decomposition still gives the right answer which can be covariantised
to the full answer in these cases also.

For example, if we reconsider the 6 point $k=2$ example we looked at
in section~\ref{sec:six-point-nnmhv}, we can use the residual $GL(6)$
symmetry to set 5, ($a,b,c,d,e$), of the 8 variables   to unity:
\begin{align}
Y=\left(
\begin{array}{cccccc}
 1 & 1 & 1 & 0 & 1 & 1 \\
 0 & 1 & f & 1 & g & h \\
\end{array}
\right)\ .
\end{align}

Implementing the cylindrical decomposition procedure exactly as in
section~\ref{sec:six-point-nnmhv} we arrive at the correct answer for
the correlahedron form in these reduced variables (assuming the
measure reduces in the obvious way to $df\wedge dg \wedge dh$. The full
covariant 
form can then be obtained from this using the full symmetries.

\subsubsection{Correlahedron example}

Encouraged by the above results  we now consider the simplest non-trivial
correlation function, the 5 point NMHV correlator.

The correlahedron is the subspace $Y=Y_1\wedge..\wedge Y_6 \in \Gr(6,10)$ restricted to the region
\begin{align}
  \label{eq:23}
  \langle YX_iX_j\rangle >0 \quad i\neq j =1,..,5\ .
\end{align}

We first use $GL(10)$ to choose the 10 external points $X_{i\alpha}$
to be the basis elements. We then note that there is a residual $GL(2)^5
\subset GL(10)$ which
leaves this external data fixed (up to the $GL(2)$ acting on each $X_{i\alpha}$).

Thus we have a $Y \in \Gr(6,10)$ with  a $GL(k)$ symmetry acting on
the left and a $GL(2)^5$ on the right. 
We can put coordinates on this as follows
\begin{align}
  \label{eq:24}
Y=
\left(  \begin{array}{c}
    Y_1\\\vdots\\Y_6
  \end{array}
\right)
=  \left(
\begin{array}{cccccccccc}
 1 & 0 & 0 & 0 & 0 & 0 & 1 & 0 & 1 & 0 \\
 0 & 1 & 0 & 0 & 0 & 0 & 0 & 1 & 0 & 1 \\
 0 & 0 & 1 & 0 & 0 & 0 & 1 & 0 & a & 0 \\
 0 & 0 & 0 & 1 & 0 & 0 & 0 & 1 & 0 & b \\
 0 & 0 & 0 & 0 & 1 & 0 & 1 & 0 & c & 1 \\
 0 & 0 & 0 & 0 & 0 & 1 & 0 & 1 & e & f \\
\end{array}
\right)
\end{align}
We claim it is always possible to put $Y$ in this form using the above
symmetries. First use  the $GL(6)$ on the left to set the matrix
consisting of the first six columns of $Y$ to the identity.
Then use the   residual
$GL(2)^5$ 
acting on the right of $Y$, together with compensating $GL(2)^3\subset
GL(6)$ acting on the left to
restore the form of $Y$. We can use this residual $GL(2)^5$ to fix the
final four columns of $Y$ to the above form. For example, in the last
two stages after fixing all but the bottom right 4$\times$2
block there is still a residual symmetry, block diagonal 
diag$(G,G,G)\subset GL(6)$ on the left and
diag$(G^{-1},G^{-1},G^{-1},G^{-1},G^{-1})$ on the 
right, for $G$ a $GL(2)$ matrix, leaving all but the bottom right
4$\times$2 block invariant. Using this we can diagonalise the $a,b$
2$\times2$ matrix and set one of the off diagonal components of the
bottom 2$\times2$ bocks to 1. 
The only remaining symmetry is
 a matrix on the left proportional to the
identity and also on the right (with the inverse  factor). This
$GL(1)$ does
not and can never act on $Y$. Note that the number of
variables of $Y$ is the dimension of  $Y$ (60) minus the dimension of
the residual
symmetry  $GL(6) \times
GL(2)^5/GL(1)$ $(6^2+5.2^2-1=55)$, giving 5, in agreement with our
five remaining variables $a,b,c,e,f$.

Having reduced the variables down to the minimal number consistent
with the symmetries,  we now perform the cylindrical decomposition
procedure.
The region~\eqref{eq:23} then corresponds to the restrictions
\begin{align}
  \label{eq:25}
- e + c f > 0, a b > 0, a b - b c -  e - a f + c f > 0, 
 1 - c -  e - f + c f > 0, (-1 + a) (-1 + b) > 0
\end{align}
which upon rewriting as a cylindrical decomposition and converting to
a differential form according to~\eqref{eq:15} gives
\begin{align}
  \label{eq:26}
  \frac{(a-b)^2 da\,db\,dc\,de\,df}{(a-1) a (b-1) b ( e-c f) (c
  (-f)+c+ e+f-1) (a b-a f-b c+c f- e)}\ .
\end{align}

Let us then compare this with the 
 expected answer for the correlator~\eqref{eq:3}. With the choice of
 variables for $Y$~(\ref{eq:24}) this gives 
\begin{align}
  \label{eq:27}
  \frac{   d\mu(a,b,c,e,f)}{(a-1) a (b-1) b ( e-c f) (c (-f)+c+ e+f-1) (a b-a
  f-b c+c f-e)}\ 
\end{align}
where $d\mu(a,b,c,e,f)$ is the measure, $\langle Y d^4Y_1\rangle \dots \langle Y
  d^4Y_6\rangle$ reduced to these variables. Remarkably, we get
  complete agreement 
  on identifying $d\mu(a,b,c,e,f)=(a-b)^2 da\,db\,dc\,de\,df$. Note
  that the term $(a-b)^2$ is indeed the natural measure factor, the 
Vandermonde determinant squared, one
  obtains 
  when writing an integral measure on $GL(2)$ invariant under
  conjugation in terms of its eigenvalues. Here it was produced
  directly by the cylindrical decomposition procedure.

So we see that in this case at least, the cylindrical decomposition
procedure still works, once all symmetries are correctly taken into
account. We leave it to future investigations to firm up this proposal.

\section{Conclusions}
\label{sec:conclusions}

In this paper we have presented the definition of a new
geometric object,  the correlahedron,  defined as a subspace of
$\Gr(n{+}k,n{+}k{+}4)$.  We have provided much evidence for its equivalence  to the correlator of stress-energy multiplets
$G_{n;k}$. We have shown how to obtain the
volume form 
associated with the 
squared amplituhedron region and its equivalence to squared
amplitude expressions in a number of examples. In
the process we developed a simple algorithmic procedure for finding the
volume form from the region.  We have also  shown 
that the correlahedron as a geometric region reduces to the 
squared amplituhedron as a geometric region (in fact many different
squared amplituhedrons in general)  via a geometric procedure of
freezing the space to a certain boundary and then projecting. The
exact same reduction procedure, applied to all known correlator
expressions (recast as volume forms in correlahedron space)  reduces
them correctly to the corresponding squared amplitude expressions.    

We believe this gives substantial evidence that the correlahedron
geometry is equivalent to the correlator. However the extraction of
the relevant volume form is more problematic for the correlator than
for the squared amplitude both computationally and conceptually.
We overcome both problems in the simplest possible example, by
exhausting the full additional symmetries of the problem and only then 
implementing the cylindrical decomposition procedure. Clearly more
work needs to be done however to make the procedure fully concrete, in
particular a fuller understanding of the reduced measure.

Our work leaves room for a number of other directions to pursue from
here. One of the many remarkable aspects of the -hedron programme,
before one even considers the geometric one is
the bosonisation of nilpotent
invariants. This provides an entirely new way to explore nilpotent
superconformal invariants in a completely bosonic framework as we saw
in section~\ref{sec:expl-corr-form} and we
believe this aspect alone deserves further investigation. One
pertinent technical question here is how to extract explicit
component correlation functions directly from this bosonised form.

The maximally nilpotent correlator, which in the lightlike limit
leads to a sum of products of amplitudes with their conjugates,
is a simpler object than the  amplitudes themselves and
indeed recent high loop  four- and five-loop amplitude expressions
have been calculated via 
the
correlator~\cite{Ambrosio:2013pba,Bourjaily:2015bpz,Bourjaily:2016evz}. It
would be interesting to understand the extent to which one can  extract
the separate amplitude expressions  from the maximally nilpotent
correlator  at higher than five points.

Another recent development is the computation of higher loop
correlators of 
higher charge BPS operators~\cite{Chicherin:2015edu}. It would be
interesting to explore how/whether the correlahedron 
generalises to yield these.

In a different  direction, it is important to find a systematic proof of the equivalence between 
 'hedra and amplitudes and correlators. One
approach is to directly work at the level of  twistor Feynman
graphs.  These can individually be mapped to regions in hedron space that together provide a tessellation of the hedron. This seems to
be problematic for the amplituhedron itself where sign ambiguities seem to lead to an obstruction to the programme for even $k$.  There is some hope that the more explicit definition provided by the squared amplituhedron might remove these obstructions.  Nevertheless, unlike the the tesselations provided by the BCFW terms represented in the positive grassmannian, individual `tiles' seem to have to lie both inside and outside the hedron.  As it stands, however, the BCFW description does not apply to correlators, so at this stage, there doesn't seem to be an alternative to the twistor space Feynman diagrams.

\vspace{-0pt}\section*{Acknowledgments}\vspace{-6pt}
The authors gratefully acknowledge helpful discussions and initial
collaboration with Reza Doobary and helpful discussion with Jaroslav
Trnka and Nima Arkani-Hamed.  BE is supported by the German research
foundation, DFG "eigene Stelle" ED 78/4-3. PH is supported by  an
STFC Consolidated Grant ST/L000407/1 and the Marie Curie network GATIS
(gatis.desy.eu) of the European Union's Seventh Framework Programme
FP7/2007-2013 under REA Grant Agreement No.\ 317089 (PH). LJM is
supported by EPSRC grant EP/M018911/1.

\appendix

\section*{APPENDIX}

\section{Lightlike limit of NMHV six points \texorpdfstring{$G_{6;1}\rightarrow
  \cA_{6;1}$}{G61->A61} }
\label{sec:lightlike-limit-nmhv}

\subsection{Maximal lightlike limit}

As a highly non-trivial example of this lightlike limit procedure 
we here explicitly reduce the six point ``NMHV'' correlator $G_{6;1}$
found in~\cite{Chicherin:2015bza} to the NMHV 6-point amplitude by performing the
``freeze and project'' procedure outlined for the correlahedron form in
section~\ref{sec:lightl-limit-corr}. In particular we implement the
reduction in the form~\eqref{eq:33} and we do it in terms of specific
coordinates rather than covariantly.

The correlahedron form for this case is given explicitly
in~\eqref{ABC} where $Y\in \Gr(7,11)$. We choose our basis  
to be $\{Y_1,\dots,Y_6,X_{12},X_{22},\dots X_{52}\}$ where $Y_1,
\dots, Y_6$ are frozen to $Y_i=X_{i1} - X_{i{+}1\,2} (i=1\dots 6)$ as
in~\eqref{eq:33}. The projection then projects out the first 6
coordinates in this basis and projects onto the final 5 coordinates. 
So with respect to  this basis we have
\begin{align}
   X_{i2}^\cA  &=-Z_i^\cA = -\delta^\cA_{i+6}\qquad   &X_{i1}^\cA
  &=X_{i{-}1\,2}^\cA-Y_{i{-}1}^\cA = -\delta^\cA_{i+5}-\delta^\cA_{i-1}  \quad i=1..5  \notag\\
X_{62}^\cA&=-Z_6^\cA=(A,B,\dots,J,1) \qquad
                                               &X_{11}^\cA&=X_{62}^\cA-Y_6^\cA=(A,..,E,F{-}1,G,..J,1)\
      \notag\\
Y_i^\cA&=\delta_i^\cA                                          &Y_7^\cA&=(0..0,1,a,b,c,d)            \ .
\end{align}
The projection operation then corresponds to projecting onto the last
5 coordinates. In particular  we set the variables
$A,..,F\rightarrow 0$. The projected points have five dimensional coordinates 
\begin{align}
  \hat Z_i  &= -\delta^{\cA'}_{i}  \quad
    i=1..5 \notag\\
\hat Z_6^{\cA'}&=-(G,..,J,1) \quad
      \notag\\
\hat Y^{\cA'}&=(1,a,b,c,d)            
\end{align}
It is straightforward (on a computer) to plug these values  into the
expression for the  correlahedron~\eqref{ABC} (see~\eqref{eq:33}). We arrive at a
  rational function of $a,b,c,d,G,H,I,J$. This rational function is
  precisely
  \begin{align}
    [12345]+[34561]+[56123]
  \end{align}
where
\begin{align}
  [ijklm]=\frac{\langle \hat Z_i\hat Z_j\hat Z_k\hat Z_l\hat Z_m
  \rangle^4}{\langle \hat Y\hat Z_i\hat Z_j\hat Z_k\hat Z_l\rangle\langle
  \hat Y\hat Z_j\hat Z_k\hat Z_l\hat Z_m\rangle\langle \hat Y\hat Z_k\hat
  Z_l\hat Z_m\hat Z_i\rangle\langle \hat Y\hat Z_l\hat Z_m\hat Z_i\hat
  Z_j\rangle\langle \hat Y\hat Z_m\hat Z_i\hat Z_j\hat Z_k\rangle}\ ,\label{eq:34}
\end{align}
which we recognise as the NMHV six-point amplituhedron form.

\subsection{Non-maximal lightlike limit}

\label{sec:five-point-lightlike}

In section~\ref{sec:non-maximal-limit-1} we performed the maximal
lightlike limit explicitly on the six point ``NMHV'' correlahedron
expression 
$G_{6;1}$. We now consider the non-maximal five-point lightlike limit
which reduces it to the five-point one-loop amplitude by implementing
the non-maximal freeze and project procedure of
section~\ref{sec:non-maximal-limit-1}.

We start with the correlahedron form given explicitly
in~\eqref{ABC} where $Y\in \Gr(7,11)$. We choose our basis  
to be $\{Y_1,\dots,Y_5,X_{61},X_{62},X_{12},X_{22},X_{32},X_{42}\}$,
where $Y_1,
\dots, Y_5$ are frozen to $Y_i=X_{i1} - X_{i{+}1\,2} (i=1\dots 5)$ as
in~\eqref{eq:33}. The projection then projects out the first 7
coordinates (the five $Y$s as well as $X_6$) in this basis and projects onto the final 4 coordinates. 

Then with respect to  this basis we have
\begin{align}
  X_{i2}^\cA  &= -Z_i^\cA = -\delta^\cA_{i+7}\qquad   &X_{i1}^\cA
  &=X_{i{-}1\,2}^\cA-Y_{i{-}1}^\cA = -\delta^\cA_{i+6}-\delta^\cA_{i-1}  \quad i=1..4  \notag\\
X_{52}^\cA&=-Z_5^\cA=(A,B,\dots,J,1) \qquad
                                               &X_{11}^\cA&=X_{52}^\cA-Y_5^\cA=(A,..,E{-}1,F,G,..J,1)\
      \notag\\
X_{61}^\cA&=\delta^\cA_{6} \qquad
                                               &X_{62}^\cA&=\delta^\cA_{7}\
      \notag\\
Y_i^\cA&=\delta_i^\cA\quad i=1..5     \notag\\
                                     Y_6^\cA&=(0..0,1,0,1,0,a,b)   & Y_7^\cA&=(0..0,1,0,1,c,d)            
\end{align}
The projection operation then corresponds to projecting onto the last
4 coordinates. In particular  we set the variables
$A,..,G\rightarrow 0$. The projected points have four dimensional coordinates 
\begin{align}
  \hat Z_i^{A'}  &= \delta^{\cA'}_{i}  \quad
    i=1..4 \notag\\
\hat Z_5^{\cA'}&=-(H,I,J,1) \quad
      \notag\\
\cL^{\cA'}_\alpha&=\left(
                   \begin{array}{cccc}
                     1&0&a&b\\
                     0&1&c&d
                   \end{array}
\right)
\end{align}

It is straightforward (on a computer) to plug these values  into the
expression for the  correlahedron~\eqref{ABC}  (see the LHS of~\eqref{eq:43}). We arrive at a
  rational function of $a,b,c,d,H,I,J$. This rational function is
  \begin{align}
    \frac{ \langle5123\rangle\langle1245\rangle}
    {\langle{\cL 12}\rangle\langle{\cL 23}\rangle\langle{\cL 51}\rangle\langle{\cL 45}\rangle}
    +\frac{ \langle1234\rangle\langle2345\rangle}
    {\langle{\cL 12}\rangle\langle{\cL 23}\rangle\langle{\cL 34}\rangle\langle{\cL 45}\rangle}
    +
    \frac{ \big( -\langle{\cL 12}\rangle \langle{ 2345}\rangle+\langle{\cL 25}\rangle\langle{ 1234}\rangle      \big)\langle1345\rangle}
    {\langle{\cL 51}\rangle\langle{\cL 12}\rangle\langle{\cL 23}\rangle\langle{\cL 34}\rangle\langle{\cL 45}\rangle}
  \end{align}
where
\begin{align}
  \langle{\cL 12}\rangle:=   \langle{\cL \hat Z_1\hat Z_2}\rangle\ .
\end{align}
This is precisely (up to a numerical factor) 
the one-loop  five-point amplitude given in~\cite{ArkaniHamed:2010gh} (eq 6.4 with $X$ chosen to be $X=45$).

\providecommand{\href}[2]{#2}\begingroup\raggedright\endgroup

\end{document}